\documentclass[a4paper,11pt]{article}

\usepackage[cmex10]{amsmath}

\usepackage[english]{babel}
\usepackage{amscd}
\usepackage{amsfonts}
\usepackage{latexsym}
\usepackage{amsthm}
\usepackage{amssymb,amsmath}
\usepackage{enumerate}

\newtheorem{Definition}{\sc Definition}[section]
\newtheorem{Theorem}[Definition]{\sc Theorem}
\newtheorem{Lemma}[Definition]{\sc Lemma}
\newtheorem{Proposition}[Definition]{\sc Proposition}
\newtheorem{Corollary}[Definition]{\sc Corollary}

\newtheorem{Remark}[Definition]{\sc Remark}

\newtheorem{Maintheorem}[Definition]{\sc Main Theorem}

\newcommand{\BIGOP}[1]{\mathop{\mathchoice

{\raise-0.22em\hbox{\huge $#1$}}

{\raise-0.05em\hbox{\LARGE $#1$}}{\hbox{\large $#1$}}{#1}}}

\newtheorem{Construction}[Definition]{\sc Construction}

\newenvironment{Proof}[1][]{\par \sc Proof{#1}. \rm\ignorespaces}{\hspace*{\fill}$\triangle$\vspace{1ex}\ignorespacesafterend}

\newcommand{\R}{\mathbb{R}}

\newcommand{\Z}{\mathbb{Z}}
\newcommand{\N}{\mathbb{N}}
\newcommand{\F}{\mathbb{F}}
\newcommand{\Fq}{\F_q}

\newcommand{\E}{\mathbb{E}}

\newcommand{\EE}{\mathcal{E}}

\newcommand{\NN}{\mathcal{N}}

\newcommand{\AAA}{\mathcal{A}}
\newcommand{\PP}{\mathcal{P}}
\newcommand{\HH}{\mathcal{H}}

\newcommand{\CC}{\mathcal{C}}

\newcommand{\UU}{\mathcal{U}}

\newcommand{\charf}{\operatorname{char}}

\newcommand{\rk}{\operatorname{rk}}

\newcommand{\Quad}{\operatorname{Quad}}
\newcommand{\ev}{\operatorname{ev}}
\newcommand{\Rad}{\operatorname{Rad}}
\newcommand{\codim}{\operatorname{codim}}

\newcommand{\gbinom}[2]{\genfrac{[}{]}{0pt}{}{#1}{#2}}

\begin{document}
%
\title{Squares of Random Linear Codes}
%
%
%
\author{Ignacio Cascudo\thanks{Aarhus University, Denmark. Email:  \texttt{ignacio@cs.au.dk}.}, Ronald Cramer\thanks{CWI Amsterdam and the Mathematical Institute, Leiden University, The Netherlands.
Email: \texttt{cramer@cwi.nl, cramer@math.leidenuniv.nl}.}, Diego Mirandola\thanks{CWI Amsterdam and Mathematical Institute, Leiden University, The Netherlands, and Institut de Math\'ematiques de Bordeaux, UMR 5251, Universit\'e de Bordeaux, France. Email: \texttt{diego@cwi.nl}.} and Gilles Z\'emor\thanks{Institut de Math\'ematiques de Bordeaux, UMR 5251, Universit\'e de Bordeaux, France. Email: \texttt{zemor@math.u-bordeaux.fr}.}
}

\maketitle
{\let\thefootnote\relax\footnotetext{I.~Cascudo acknowledges support from the Danish National Research Foundation and The National Science Foundation of China (under the grant 61361136003) for
the Sino-Danish Center for the Theory of Interactive Computation and from the Center for Research in Foundations of Electronic Markets (CFEM), supported by the Danish
Strategic Research Council. Moreover, the research was partially carried out while I.~Cascudo was at CWI Amsterdam, The Netherlands, supported by STW Sentinels program under Project 10532 and by Cramer's NWO VICI Grant “Mathematics of Secure Computation”. D.~Mirandola acknowledges the support of the ALGANT-DOC programme.
}}
{\let\thefootnote\relax\footnotetext{Material in this paper was presented at the ``Mathematics of Information-Theoretic Cryptography'' workshop, Leiden, May~13-25~2013 and at the ``Algebra, Codes and Networks'' workshop, Bordeaux, June~16-20~2014.
To appear on IEEE Transactions on Information Theory.}}
\begin{abstract}
Given a linear code $C$, one can define the $d$-th power of $C$ as the span of all componentwise products of $d$ elements
 of $C$. A power of $C$ may quickly fill the whole space. Our purpose is to answer the following question: does the
square of a code ``typically'' fill the whole space? We give a positive answer, for codes of dimension $k$ and length roughly
$\frac{1}{2}k^2$ or smaller.
Moreover, the convergence speed is exponential if the difference $k(k+1)/2-n$ is at least linear in $k$.
The proof uses random coding and combinatorial arguments, together with algebraic tools
involving the precise computation of the number of quadratic forms of a given rank, and the number of their zeros.
\end{abstract}

%
\title

\section{Introduction}
Let $K$ be a field and denote by $*$ the coordinatewise
product of vectors of $K^n$, so that:
$$(x_1,\dots,x_n)*(y_1,\dots,y_n) = (x_1y_1,\dots,x_ny_n).$$
When $V$ and $W$ are subspaces of $K^n$ let us denote similarly
by $V*W$ the subspace generated by all $*$-products of vectors of $V$
and $W$, i.e. $V*W:=\langle x*y:x\in V,y\in W\rangle$. We also use the
shorthand $V^{*1}=V$, $V^{*2} := V*V$ and define inductively the powers of $V$,
$V^{*d}:=V*V^{*(d-1)}$ for $d>1$.

When $K=\F_q$ is a finite field and $C$ is a $q$-ary linear code, asking what
are the possible parameters of the linear code $C^{*2}$ arises in a
number of different contexts and has attracted a lot of attention
recently.
Possibly one of the earliest appearances of this question in coding
theory goes back to \cite{Pel92} where it is relevant to the notion
of  error-locating pairs used for algebraic decoding.

``Products'' and ``squares'' of codes are the primary focus of 
work on secret sharing~\cite{CC06,CCCX09,CCX11,CCX12} and its application to secure multi-party computation~\cite{CDM00}.
To share a secret vector $s\in\Fq^k$ among $n$ players using a
linear code $C\subseteq\Fq^{n+k}$, one standardly chooses a random codeword with some fixed $k$-tuple of coordinates equal to $s$: the other coordinates are the shares.
When two secrets $s$ and $t$ are shared in this way, summing
coordinatewise the share vectors gives naturally a share vector of the
coordinatewise sum $s+t$ of the secrets. When one considers the $*$-product of the
share vectors, one obtains a share of the product $s*t$, but for a
different secret-sharing scheme, namely that associated to the
$*$-product code $C^{*2}$.
Since the parameters of a code are relevant
to the associated secret-sharing scheme, studying the parameters of
$C^{*2}$ becomes important.
More precisely, interest is focused on families of linear codes ${(C_i)}_{i\in\N}$ of unbounded length, such that the families of the dual codes ${(C_i^\perp)}_{i\in\N}$ and of the squares ${(C_i^{*2})}_{i\in\N}$ are asymptotically good.
A family of codes satisfying this property yields linear secret-sharing schemes on arbitrarily many players with good parameters (privacy, reconstruction, multiplication)~\cite{CCCX09}.
Such families were first constructed, over almost all finite fields, in~\cite{CC06} using techniques from algebraic geometry (asymptotically good towers of algebraic function fields).
This work was subsequently extended in~\cite{CCCX09,CCX11} involving novel algebraic-geometric ideas.
We remark that no elementary construction is known so far.

Secret sharing has as main motivation and application secure multi-party computation (MPC).
Any linear secret-sharing scheme yields an MPC protocol~\cite{CDM00}, and the family of all malicious coalitions of players the protocol can tolerate depends on the parameters of the LSSS listed above.

Besides its original application, the result of~\cite{CC06} played a central role in the paper~\cite{IKOS09} on the ``secure MPC in the head'' paradigm: here secure MPC is used as an abstract primitive for efficient two-party cryptography.\footnote{For an extensive treatment of the interplay between secure multiparty computation, (arithmetic) secret sharing, codes and algebraic geometry, please consult~\cite{mpc_book}.}
Among other subsequent fundamental results, let us mention that
asymptotically good codes whose dual and square are also asymptotically good
are an essential ingredient in the recent constructions of efficient
unconditionally secure oblivious transfer protocols from noisy
channels~\cite{IKOPSW11,OZ}.

The same issue is also pertinent to algebraic complexity theory: there
one wishes to express multiplication in the extension field $\F_{q^m}$
through a bilinear algorithm involving a small number of multiplications
in $\F_q$, see \cite{BP11,CCXY11,Ran12,CCX14} for recent developments.

Motivated in part by these applications, asymptotically good codes whose squares are also asymptotically good (and we impose no conditions on the duals) have been shown to exist for all finite fields in~\cite{Randriam13}. This construction carefully combines algebraic geometric codes that have asymptotically good higher powers, which can be constructed over  large enough finite fields, with a field descent concatenation technique. Again, no elementary construction is known in this case.

Powers of linear codes also turn up in lattice constructions, as was
recently elaborated on in \cite{KO13}. If $C$ is a binary linear code,
then, abusing notation by identifying $C$ with its natural lift in
$\Z^n$, the most natural lattice construction from $C$ is $\Lambda =
C+2\Z^n$
(construction A in Conway and Sloane's terminology \cite{CS99}). The
minimum Euclidean norm of a lattice vector is then
$\min(\sqrt{d_{\min}(C)},2)$, where $d_{\min}(C)$ is the minimum Hamming distance of
the code $C$. If one wishes to generate from the code $C$ a lattice with larger
Euclidean distance, one may try to construct the lattice generated by
$C+4\Z^n$: a close look shows that this lattice actually equals
$$C + 2C^{*2} + 4\Z^n$$
and its minimum Euclidean norm is
$$\min\left(\sqrt{d_{\min}(C)},2\sqrt{d_{\min}(C^{*2})}, 4\right).$$
One may generalize the construction to $C+2C^{*^2} + 4C^{*4} + 8\Z^n$
and so on, or more generally to (construction D \cite{CS99})
$C_0+2C_1+\cdots + 2^{\ell-1}C_{\ell-1}+2^{\ell}\Z^n$, which is a
lattice if and only if $C_j^{*2}\subset C_{j+1}$, a fact not usually
explicitely stated in the literature.

Finally, there has been some recent use of $*$-squares in the 
cryptanalysis of variants of the McEliece cryptosystem
\cite{Faugere2011,CGGOT13,COT13,COT14}.
The idea that is exploited is that Goppa codes have a $*$-square that
has a substantially smaller dimension than typical random linear
codes: this allows to build a distinguisher which can be used to attack the cryptosystem.

The motivation for a systematic code-theoretic study of $*$-squares is
therefore quite strong.
For a wide collection of results on the topic see~\cite{Randriam_draft} and references therein.
With a view to contribute to such an endeavour,
our concern in the present work is with the
dimension of squares of random linear codes: we see that this is especially
relevant in particular to the last application to cryptanalysis.

Since a generating set of vectors for the square of a code $C$ of
dimension $k$ can be constructed by taking
all possible $k(k+1)/2$ products of two elements of a basis of the
code $C$, it is reasonable to expect that a randomly chosen code of block
length $n < k(k+1)/2$ has a $*$-square which fills up the whole
space, i.e. $C^{*2}=\F_q^n$. However, linear relations between
products of elements of $C$ are not typically independent random
events, and one has to overcome a certain number of obstacles to
prove such a statement. Our main result is indeed to show that
when the difference $k(k+1)/2-n$ goes to infinity as a function of
$k$, however slowly, the probability that a random code of length $n$
and dimension $k$ has a square different from $\F_q^n$ goes to zero.
We also study the speed of convergence, which is exponential if the difference $k(k+1)/2-n$ is at least linear in $k$, and the limiting case
$n=k(k+1)/2$. We shall also consider the slightly easier case when the blocklength $n$
is such that $n\geq k(k+1)/2$: 
we obtain that with probability tending to $1$ when $n-k(k+1)/2$ goes
to infinity, the dimension of the square of the random code is exactly
$k(k+1)/2$. Again, this convergence is exponentially fast if $n-k(k+1)/2$ is at least linear in $k$.
Previously, the best-known fact on this problem was given
by Faug\`ere et al. in \cite{Faugere2011} who proved that for $n\geq
k(k+1)/2$ and for any function $\omega(k)$ that goes to infinity
with $k$, the dimension of the square of the random code is at least
$k(k+1)/2 - k\omega(k)$ with probability tending to $1$ when $k$ goes
to infinity.

Our techniques break significantly with the approach of
\cite{Faugere2011} and combine the study of the dual
distance of the square of a random code, and the distribution of
zeros of random quadratic forms. In the next section we describe our
results precisely and give an overview of our proofs and the structure
of the paper.

\section{Overview}\label{sec:statement}
Throughout this paper, $q$ denotes a fixed prime power and $\Fq$ a field with $q$ elements.

We first define the probabilistic model we shall work with.
For all positive integers $n\geq k$, we define $\CC(n,k)$ to be the
family of all $[n,k]$-codes over $\Fq$ whose 
first $k$ coordinates make up an information set: equivalently,
members of $\CC(n,k)$ have a
generator matrix which can be written in systematic form, i.e.\ as
$$
G=
\left(
\begin{matrix}
1&&\\
&\ddots&\\
&&1
\end{matrix}\quad
\vline\quad
\begin{matrix}
&&\\
&A&\\
&&
\end{matrix}\quad
\right),
$$
for some $k\times(n-k)$ matrix $A$. We endow $\CC(n,k)$ with the uniform distribution.
Since codes of $\CC(n,k)$ are 
in one-to-one correspondence with $k\times (n-k)$ matrices $A$, choosing
a random element of $\CC(n,k)$ amounts to choosing a random uniform matrix $A$.

\begin{Remark}\label{rem:prob_model}
There are several possible choices for the probabilistic
model. An alternative way of choosing a random code consists of
choosing its generator matrix uniformly at random among all $k\times n$ matrices. Yet another alternative is to consider
the uniform distribution among all codes of length $n$ and dimension
$k$. The first alternative probability distribution has the disadvantage that the resulting code may be of dimension
$<k$. The second alternative distribution is perhaps the most
theoretically elegant but makes it somewhat cumbersome to use the puncturing
arguments that we will work with, hence the above choice of a
probabilistic model. In Section~\ref{sec:prob_model} we
shall argue however that our results are not altered significantly
under these alternative probability distributions.
\end{Remark}


Our main result is:

\begin{Maintheorem}\label{mthm:main}
Let $n\colon\N\to\N$ be such that $k(k+1)/2\geq n(k)\geq k$ for all $k\in\N$ and define $t\colon\N\to\N,t(k):=k(k+1)/2-n(k)$.
Then there exist constants $\gamma,\delta\in\R_{>0}$ such that, for all large enough $k$,
$$
\Pr(C^{*2}=\Fq^{n(k)})\geq 1-2^{-\gamma k}-2^{-\delta t(k)},
$$
where $C$ is chosen uniformly at random from $\CC(n(k),k)$.
\end{Maintheorem}


For lengths $n$ that are larger than $k(k+1)/2$, we also have:

\begin{Theorem}\label{thm:biglength}
Let $n\colon\N\to\N$ be such that $n(k)\geq k(k+1)/2$ for all $k\in\N$ and define $s\colon\N\to\N,s(k):=n(k)-k(k+1)/2$.
Then there exists a constant $\hat \delta\in\R_{>0}$ such that, for all large enough $k$,
$$
\Pr\left(\dim C^{*2}=\frac{k(k+1)}{2}\right)\geq 1-2^{-\hat\delta s(k)},
$$
where $C$ is chosen uniformly at random from $\CC(n(k),k)$.
\end{Theorem}

Strangely enough, Theorems~\ref{mthm:main} and \ref{thm:biglength} are
not quite symmetrical. In particular the term~$2^{-\gamma k}$ is
absent from the statement of Theorem~\ref{thm:biglength} but can not
be avoided in Theorem~\ref{mthm:main}: this is because with
probability at least $1/q^k$, the random matrix $G$ will contain a
column of zeros, or two identical columns, in which case the square
$C^{*2}$ can not be equal to $\Fq^{n(k)}$.
The two theorems will not require exactly the same
methods and Theorem~\ref{mthm:main} will need more work
than Theorem~\ref{thm:biglength}. We shall deal with them separately.

Our first step towards establishing Theorem~\ref{mthm:main} will be to
estimate the expected minimum distance of the dual of the square of a
random code of length $k(k+1)/2$. Specifically, we shall prove:

\begin{Proposition}\label{prop:distdualsq}
There exist constants (depending only on $q$) $c,\widetilde{c}\in\R_{>0}$ such that, for all large enough $k$,
if $C$ is chosen uniformly at random from $\CC(k(k+1)/2,k)$ then
$$
\Pr\left(d_{\min}({(C^{*2})}^\perp)\leq c\cdot \frac{k(k+1)}{2}\right) \leq 2^{-\widetilde{c}k}.$$
\end{Proposition}

This last proposition enables us to use puncturing arguments.
In our probabilistic model, a random code of length $n$ can be
obtained by first choosing a random code of length $n+t$ and then
puncturing $t$ times on a random position. The probability that a
punctured code has the same dimension as the original code is
well-separated from zero whenever the dual distance of the original
code is large enough. This fact 
will be enough in itself to establish
the following weaker version of Main Theorem~\ref{mthm:main}.

\begin{Theorem}\label{thm:weak}
There exist constants (depending only on $q$) $c,\widetilde
c\in\R_{>0}$ such that, if $n\colon\N\to\N$ 
satisfies
$$k\leq n(k)\leq c\cdot\frac{k(k+1)}{2}$$ for all $k\in\N$
then, for all large enough $k$,
$$\Pr(C^{*2}=\Fq^{n(k)})\geq 1-2^{-\widetilde c k},$$ 
where $C$ is chosen uniformly at random from $\CC(n(k),k)$.
\end{Theorem}

However, in order to deal with block lengths that approach the upper
bound
$k(k+1)/2$ on the dimension of the square of $C$, and prove the full-fledged
Main Theorem~\ref{mthm:main}, we need some additional ingredients.
 
Given an $[n,k]$-code $C$ and denoting by $\pi_1,\dots,\pi_n\in\Fq^k$ the columns of a generator matrix of $C$, define the linear map
$$
\begin{matrix}
\ev_C\colon&\Quad(\Fq^k)&\to&\F_q^n,\\
&Q&\mapsto&(Q(\pi_1),\dots,Q(\pi_n))
\end{matrix}
$$
where $\Quad(\Fq^k)$ denotes the vector space of quadratic forms on $\Fq^k$. Then one can see that the image of $\ev_C$ does not depend on the choice of a generator matrix of $C$, and it is equal to $C^{*2}$, see~\cite[\S 1.31]{Randriam_draft}. In particular, $C^{*2}=\Fq^n$ if and only if $\ev_C$ is surjective.
Moreover, by basic linear algebra $C^{*2}=\Fq^n$ if and only if
$$
\dim\ker\ev_C=\dim\Quad(\Fq^k)-n=\frac{k(k+1)}{2}-n.
$$
So it makes sense to focus on this kernel. We view its cardinality as
a random variable, with distribution induced by the uniform
distribution of $C$ over $\CC(n,k)$: formally, for all positive integers $n\geq k$ we define
$$
X(n,k):=|\ker\ev_C|.
$$

Our main intermediate result, of interest in its own right, is:

\begin{Theorem}\label{thm:expectation}
We have that
$$
\lim_{k\to\infty}\E\left[X\left(\frac{k(k+1)}{2},k\right)\right]=2.
$$
\end{Theorem}

A simple use of Markov's inequality will then give us that, for a
random code $C$ of length $k(k+1)/2$, the
probability that the codimension of $C^{*2}$ does not exceed $\ell$,
$$\Pr\left(\dim C^{*2} \geq \frac{k(k+1)}{2}-\ell\right)$$
tends to $1$ when $\ell$ goes to infinity, furthermore exponentially
fast if $\ell$ is linear in $k$. Puncturing arguments, again relying
on Proposition~\ref{prop:distdualsq}, will enable us to conclude the
proof of Theorem~\ref{mthm:main} when the block length $n$ is well
separated from $k(k+1)/2$.

As a by-product, Theorem~\ref{thm:expectation} also enables us to deal easily with
the case when $n\geq k(k+1)/2$. Theorem~\ref{thm:biglength} will
follow as a straightforward consequence.


We conclude this overview by giving a rough idea of the proof of
Theorem~\ref{thm:expectation}. It involves computing the number of
zeros of a quadratic form of given 
rank and the number of quadratic forms of given rank; the results we
need are stated precisely in Section~\ref{sec:quadforms} and a
detailed proof is provided in the Appendix.

By definition, for all positive integers $m\geq k$ we have
\begin{align*}
\E[&X(m,k)]=\\
&=\E[|\{Q\in\Quad(\Fq^k):Q(\pi_1)=\dots=Q(\pi_m)=0\}|],
\end{align*}
where we can assume that, for $i=1,\dots,k$, $\pi_i=e_i$ is the $i$-th
unit vector  while
$\pi_{k+1},\dots,\pi_m\in\Fq^k$ have independent, uniform distribution
over $\Fq^k$, by definition of the family $\CC(m,k)$ and our
probabilistic model.

Note that the conditions $Q(e_1)=\dots=Q(e_k)=0$ are independent (in the sense of linear algebra), hence the subspace
$$
S:=\{Q\in\Quad(\Fq^k):Q(e_1)=\dots=Q(e_k)=0\}
$$
of $\Quad(\Fq^k)$ has dimension $k(k-1)/2$.
Moreover, as $\pi_{k+1},\dots,\pi_m\in\Fq^k$ are independent (in the sense of probability), we have
\begin{align*}
\Pr(Q(\pi_{k+1})&=\dots=Q(\pi_m)=0)=\\
&={\Pr(Q(\pi_{k+1})=0)}^{m-k}={\left(\frac{|Z(Q)|}{q^k}\right)}^{m-k}
\end{align*}
for any $Q\in\Quad(\Fq^k)$. Here $Z(Q)$ denotes the zero set of $Q$.
Finally, by linearity of the expectation we have
\begin{align}
\E[&X(m,k)]=\nonumber\\
&=\E[|\{Q\in S:Q(\pi_{k+1})=\dots=Q(\pi_m)=0\}|]=\nonumber\\
&=\sum_{Q\in S}{\left(\frac{|Z(Q)|}{q^k}\right)}^{m-k}.\label{eq0}
\end{align}

Now if it were true (it is not) that all non-zero quadratic forms on
$\Fq^k$ have $q^{k-1}$ zeros, we would have, when we set $m=k(k+1)/2$,
$$
\E[X(m,k)]=1+\frac{1}{q^{m-k}}(q^\frac{k(k-1)}{2}-1)\longrightarrow2
$$
 ``proving'' Theorem~\ref{thm:expectation}.
However, even though it is false that all non-zero quadratic forms on $\Fq^k$ have $q^{k-1}$ zeros, this still holds ``on average'':
roughly speaking, most quadratic forms have $q^{k-1}$ zeros, quadratic forms whose number of zeros is far from this value are those of small rank, and the number of such forms is so small that it contributes almost nothing to the expectation.
In other words, the expectation behaves as if it were true that all
non zero quadratic forms on $\Fq^k$ have $q^{k-1}$ zeros.

The rest of the paper is organized as follows. Section
\ref{sec:weak} is devoted to proving Proposition~\ref{prop:distdualsq} and
Theorem~\ref{thm:weak}.
Section \ref{sec:quadforms} states the results that we need on
quadratic forms, namely the number of forms of a given rank, and the
number of their zeros. Some of these results can be found in the
literature, but only in part, and we have felt it useful to derive
what we need in a unified way: this is provided in the Appendix so as
not to disrupt the flow of the paper.
Finally, in Section \ref{sec:proof} we use the results of Section
\ref{sec:quadforms} to derive Theorem~\ref{thm:expectation}. 
Theorem~\ref{thm:biglength} is then derived as an almost immediate consequence.
We then
apply the methods and results of Section~\ref{sec:weak} to conclude
the proof of Theorem~\ref{mthm:main}.

\section{Proof of Theorem~\ref{thm:weak}}\label{sec:weak}
In this section we prove Proposition~\ref{prop:distdualsq} and Theorem~\ref{thm:weak}, the weaker version of our main result.
We start by introducing some notation and classical results that we shall need.

\begin{Definition}[Gaussian binomial coefficient]\label{def:gaussian}
For all non-negative integers $n\geq k$, we define the $q$-ary Gaussian binomial coefficient to be
$$
{\gbinom{n}{k}}_q:=
\prod_{i=1}^k\frac{q^{n-k+i}-1}{q^i-1}.
$$
\end{Definition}

By convention, we define a product with no factors to be equal to $1$. This is the case if $k=0$.
As $q$ is assumed to be fixed, it will be suppressed from the notation from here on.
It is well-known that the Gaussian binomial coefficient $\gbinom{n}{k}$ equals the number of $k$-dimensional subspaces of any $\Fq$-vector space of dimension $n$.


\begin{Remark}\label{remark:gaussianbound}
For all non-negative integers $n\geq k$, we bound
$$
\gbinom{n}{k}\leq 2^kq^{k(n-k)}.
$$
This holds as $\gbinom{n}{k}$ is the product of $k$ terms, and each term is bounded by $2q^{n-k}$.
\end{Remark}

\begin{Definition}[entropy function]
The $q$-ary entropy function is defined by
$$
H_q(x):=x\log_q(q-1)-x\log_qx-(1-x)\log_q(1-x)
$$
for all $0<x\leq 1-q^{-1}$.
\end{Definition}

Again, from here on $q$ will be suppressed from the notation. In
particular, all logarithms will be in base $q$. 
The following lemma is folklore, see e.g.~\cite[\S 2.10.3]{Huffman} for a proof.

\begin{Lemma}\label{lemma:entropy}
For all $0<\delta\leq 1-q^{-1}$ and all integers $n$, we have
$$
\sum_{i=0}^{\lfloor\delta n\rfloor}\binom{n}{i}{(q-1)}^i\leq q^{nH(\delta)}.
$$
\end{Lemma}
%
%

For ease of notation, we define $m\colon\N\to\N$ by $m(k):=k(k+1)/2$.
Also, recall that, given a code $C$, we denote by $C^\perp$ its dual and by $d_{\min}(C)$ its minimum distance.

We prove now Proposition~\ref{prop:distdualsq}. 



\begin{Proof} [ of Proposition~\ref{prop:distdualsq}]
Let $C\in\CC(m(k),k)$.
By definition, $C$ admits a generator matrix of the form
$$
\left(
\begin{matrix}
1&&\\
&\ddots&\\
&&1
\end{matrix}\quad
\vline\quad
\begin{matrix}
&g_1&\\
&\vdots&\\
&g_k&
\end{matrix}\quad
\right).
$$
Note that a uniform random selection of $C$ from $\CC(m(k),k)$ induces an independent, uniform random selection of $g_1,\dots,g_k$ from $\Fq^{m(k)-k}$.
We consider the code
$$
\langle g_i*g_j:1\leq i\leq k/2<j\leq k\rangle
$$
and we define $D$ to be its dual.
This is a code of length $k(k-1)/2$ and it is easy to see that
$$
d_{\min}({(C^{*2})}^\perp)\geq d_{\min}(D).
$$
In the following, when $D$ is involved in some probability measure, we implicitly mean that it has the distribution induced by the uniform distribution of $C$ on $\CC(m(k),k)$. We remark that this does not necessarily correspond to a uniform distribution on the set of all possible $D$'s.

For any positive integer $w$ and any code $C'$, denote by $\EE_w(C')$
the event ``there exists a non-zero codeword of $C'$ of weight
$w$''. We shall now prove the following statement, which clearly implies the Proposition.
There exist constants $c,\widetilde{c}\in\R_{>0}$ such that, for all large enough $k$,
$$
\sum_{w=1}^{cm(k)}\Pr(\EE_w(D))\leq 2^{-\widetilde{c}k}.
$$

Note that, for any positive integer $w$,
\begin{equation}\label{eq dist1}
\Pr(\EE_w(D))=\sum_{\substack{z\in\Fq^{k(k-1)/2}\\\text{of weight }w}}\Pr(z\in D).
\end{equation}
So we need to estimate, for all positive integers $w$ and all vectors $z$ of weight $w$, the probability that $z$ belongs to $D$.

We do that as follows. For $1\leq i\leq k/2$, let $x_i$ be the
projection of $g_i$ on the support of $z$. Similarly, for $k/2<j\leq
k$, let $y_j$ be the projection of $g_j$ on the support of $z$. This
defines $k$ vectors in $\Fq^w$. 
Moreover, a uniform random selection of $C$ from $\CC(m(k),k)$ induces
an independent, uniform random selection of the $x_i$'s and the
$y_j$'s from $\Fq^w$. 
Note now that if we identify $z$ with a vector of $\Fq^w$, we can
define the non-degenerate bilinear form that to any two vectors $a,b$ of $\Fq^w$
associates the quantity
$$(a|b)_z := {\mathbf 1}\cdot (z*a*b)$$
where ${\mathbf 1}$ denotes the all-one vector of $\Fq^w$ and $\cdot$
denotes the standard inner product. Let us say that $a$ and $b$ are
{\em $z$-orthogonal} if $(a|b)_z=0$. The purpose of this definition is
to note that $z\in D$ if and only if, for all $1\leq i\leq
k/2<j\leq k$, 
$x_i$ is $z$-orthogonal to $y_j$. In the computation that follows we
assume that $k$ is even, thus avoiding cumbersome floor and ceiling notation,
and giving us the same number of $x_i$'s
and of $y_j$'s, namely $k/2$. It is readily seen that the case $k$ odd
can be dealt with in a similar fashion.

For all positive integers $r<k/2$, denote by $\HH_r$ the event ``$\dim\langle x_i:1\leq i\leq k/2\rangle<r$''. Conditioning by this event, we have
\begin{align*}
\Pr(z\in D)&=\Pr(\HH_r)\Pr(z\in D|\HH_r)+\\
&+\Pr(\overline\HH_r)\Pr(z\in D|\overline\HH_r)\leq\\
&\leq\Pr(\HH_r)+\Pr(z\in D|\overline\HH_r),
\end{align*}
for any choice of $r$.
In order to estimate $\Pr(\HH_r)$, note that $\dim\langle x_i:1\leq
i\leq k/2\rangle<r$ if and only if there exists an $(r-1)$-dimensional
subspace of $\Fq^w$ containing all $x_i$'s. 
The probability that an $x_i$ falls into a given subspace of dimension
$r-1$ is $1/q^{w-r+1}$ and since the $x_i$'s are independent random
variables, the probability that all the $x_i$'s fall into the same
subspace is $1/q^{(w-r+1)k/2}$. We have therefore,
$$
\Pr(\HH_r)\leq \gbinom{w}{r-1}\frac{1}{q^{\frac{k}{2}(w-r+1)}}\leq\frac{2^{r}}{q^{(w-r)(k/2-r)}},
$$
where we have used the upper bound of Remark~\ref{remark:gaussianbound}
on the number $\gbinom{w}{r-1}$ of subspaces of dimension $r-1$.

On the other hand, $z\in D$ if and only if all $y_j$'s are $z$-orthogonal to the space $\langle x_i:1\leq i\leq k/2\rangle$,
which has dimension at least $r$, under the condition
$\overline\HH_r$. Therefore, using the independence of the random
variables $y_i$,
$$
\Pr(z\in D|\overline\HH_r)\leq\left(\frac{1}{q^r}\right)^{\frac k2}=\frac{1}{q^{\frac{rk}{2}}}.
$$
Now fixing $r:=\min\{w/2,k/4\}$ it follows that there exist two positive constants $c'$ and $c''$ such that
$$
\Pr(z\in D)\leq\frac{1}{q^{c'kw}}+\frac{1}{q^{c''k^2}}.
$$
Applying this last upper bound to \eqref{eq dist1}, we now have
\begin{align*}
\Pr(\EE_w(D))&=\sum_{\substack{z\in\Fq^{k(k-1)/2}\\\text{of weight }w}}\Pr(z\in D)\leq\\
&\leq\binom{\frac{k(k-1)}{2}}{w}{(q-1)}^w\left(\frac{1}{q^{c'kw}}+\frac{1}{q^{c''k^2}}\right)
\end{align*}
for any positive integer $w$.
Therefore, for any constant $c$ we have
\begin{align}
\sum_{w=1}^{cm(k)}\Pr(\EE_w(D))&\leq\left(\sum_{w=1}^{cm(k)} \binom{\frac{k(k-1)}{2}}{w}\frac{{(q-1)}^w}{q^{c'kw}}\right)+\nonumber\\&+\frac{1}{q^{c''k^2}}\sum_{w=1}^{cm(k)}\binom{\frac{k(k-1)}{2}}{w}{(q-1)}^w.\label{eq:bound}
\end{align}
We deal with the two terms separately.


We bound the first sum in \eqref{eq:bound} as follows,
\begin{align*}
\sum_{w=1}^{cm(k)}\binom{\frac{k(k-1)}{2}}{w}\frac{{(q-1)}^w}{q^{c'kw}}
&\leq \sum_{w=1}^{cm(k)}\left(\frac{k(k-1)}{2}\right)^w\frac{{(q-1)}^w}{q^{c'kw}}\leq\\
&\leq\sum_{w=1}^{cm(k)}q^{w(-c'k+o(k))}\leq q^{-c'k+o(k)}
\end{align*}
since there are not more than $m(k)=q^{o(k)}$ terms in the sum and
none is larger than $q^{-c'k+o(k)}$.


Writing $\binom{\frac{k(k-1)}{2}}{w}\leq\binom{m(k)}{w}$ for any $w\leq cm(k)$, the second term in \eqref{eq:bound} is upper bounded by
$$
\frac{1}{q^{c''k^2}}\sum_{w=1}^{cm(k)}\binom{m(k)}{w}{(q-1)}^w.
$$
We now set $c\leq1-q^{-1}$ and apply Lemma~\ref{lemma:entropy}:
\begin{align*}
\frac{1}{q^{c''k^2}}\sum_{w=1}^{cm(k)}\binom{m(k)}{w}{(q-1)}^w&\leq
\frac{1}{q^{c''k^2}}q^{m(k)H(c)}\leq\\
&\leq q^{(\frac{1}{2}H(c)-c'')k^2+o(k^2)}.
\end{align*}
If $c$ is such that $H(c)<2c''$ we obtain an exponentially small upper bound.
Putting everything together, we obtain
$$
\sum_{w=1}^{cm(k)}\Pr(\EE_w(D))\leq \frac{1}{q^{c'k+o(k)}}+\frac{1}{q^{\frac{1}{2}(c''-H(c)/2)k^2+o(k^2)}}
$$
and the proposition is proved.
\end{Proof}

\begin{Remark}
In the proof of the previous proposition we can take $c''=\frac{1}{8}$. Therefore the proposition holds for any
$c$ with $H(c)<1/4$. For example, for $q=2$, $c=0.041$ suffices.
\end{Remark}

We can now prove Theorem~\ref{thm:weak}.

\begin{Proof}[ of Theorem~\ref{thm:weak}]
Let $c,\widetilde c$ be the constants given by
Proposition~\ref{prop:distdualsq}. 
Let $n\colon\N\to\N$ be as in the hypothesis of the theorem.
Given $C\in\CC(n(k),k)$, we create  $V\in\CC(m(k),k)$ by adding
$m(k)-n(k)$ columns to the
systematic generator matrix of $C$. Moreover, if $C$ and all the new columns are chosen uniformly at random from $\CC(n(k),k)$ and $\Fq^k$ respectively then $V$ has the uniform distribution on $\CC(m(k),k)$. A codeword in the dual of $C^{*2}$ gives a codeword in the dual of ${V}^{*2}$ of the same weight (padding with zeros).
Hence
$$
\Pr\left(C^{*2}\not=\Fq^{n(k)}\right)\leq\Pr\left(d_{\min}({(V^{*2})}^\perp)\leq cm(k)\right)\leq 2^{-\widetilde c k}
$$
by Proposition~\ref{prop:distdualsq} and the conclusion follows.
\end{Proof}

\section{Quadratic forms}\label{sec:quadforms}


In this section we state the results that we need in the proof of our Main Theorem, as well as the definitions necessary to read such results.
For a more involved discussion, see Appendix~\ref{sec:appquadforms}, where we include full proofs of the results stated here as well.
Even though these can be found, at least partly, in the literature, we have felt it necessary to derive what we need in a unified way.

{\em Throughout this section, let $K$ be an arbitrary field.}

\begin{Definition}[quadratic form]
Let $V$ be a finite dimensional $K$-vector space.
A quadratic form on $V$ is a map $Q\colon V\to K$ such that
\begin{enumerate}[(i)]
\item $Q(\lambda x)=\lambda^2Q(x)$ for all $x\in V,\lambda\in K$,
\item the map $(x,y)\mapsto Q(x+y)-Q(x)-Q(y)$ is a bilinear form on $V$.
\end{enumerate}
The $K$-vector space of all quadratic forms on $V$ is denoted by $\Quad(V)$.
A pair $(V,Q)$ where $V$ is a finite dimensional $K$-vector space and $Q$ is a quadratic form on $V$ is called a $K$-quadratic space.
\end{Definition}

Let $(V,Q)$ be a $K$-quadratic space. With abuse of terminology, from here on we call $V$ a quadratic space, omitting the quadratic form $Q$ which defines the quadratic space structure on the vector space $V$.
We define a symmetric bilinear form $\tilde B_Q$ on $V$ by
$$
\tilde B_Q(x,y):=Q(x+y)-Q(x)-Q(y)
$$
for all $x,y\in V$. 

\begin{Definition}[radical]
The radical of the quadratic space $V$ is the $K$-vector space
$$
\Rad V:=\{x\in V:\tilde B_Q(x,y)=0\text{ for all }y\in V\}.
$$
\end{Definition}

We say that $V$ is non-degenerate (as a quadratic space) if $\tilde B_Q$ is non-degenerate (as a bilinear form), i.e.\ if $\Rad V=0$.
%

\begin{Definition}[rank]
Let $\Rad^0V:=\{x\in\Rad V:Q(x)=0\}$.
We define the rank of $Q$ to be
$$
\rk Q:=\dim V-\dim\Rad^0V.
$$
\end{Definition}

\begin{Remark}
Note that in the case $\charf K\not=2$, it holds that $Q(x)=\frac12 \tilde B_Q(x,x)$ and therefore $\Rad^0 V=\Rad V$. Hence in this case $(V,Q)$ is non-degenerate if and only if $Q$ has full rank. In the appendix we show that this is not the case if $\charf K=2$.
\end{Remark}


We are now ready to state the results we need. Theorem~\ref{thm:cor:counting-q:main} counts the number of zeros of a given quadratic form. Theorem~\ref{thm:givenrank-gen:main} counts the number of quadratic forms of a given rank.

\begin{Theorem}\label{thm:cor:counting-q:main}
Let $(V,Q)$ be an $\Fq$-quadratic space, set $k:=\dim V$ and $r:=\rk Q$.
The number of vectors $x\in V$ such that $Q(x)=0$ is
\begin{enumerate}[a.]
\item $q^{k-1}$ if $r$ is odd,
\item either $q^{k-1}-(q-1)q^{k-\frac{r}{2}-1}$ or $q^{k-1}+(q-1)q^{k-\frac{r}{2}-1}$ if $r$ is even.
\end{enumerate}
\end{Theorem}

%
%

\begin{Theorem}\label{thm:givenrank-gen:main}
For all non-negative integers $k$, the number of full-rank quadratic forms on an $\Fq$-vector space of dimension $k$ is
\begin{align*}
N(k)&=q^{\left\lfloor\frac{k}{2}\right\rfloor\left(\left\lfloor\frac{k}{2}\right\rfloor+1\right)}\prod_{i=1}^{\left\lceil\frac{k}{2}\right\rceil}(q^{2i-1}-1)=\\
&=
\begin{cases}
q^{\frac{k-1}{2}\frac{k+1}{2}}\prod_{i=1}^\frac{k+1}{2}(q^{2i-1}-1)&\text{if }k\text{ is odd,}\\
q^{\frac{k}{2}\left(\frac{k}{2}+1\right)}\prod_{i=1}^\frac{k}{2}(q^{2i-1}-1)&\text{if }k\text{ is even.}
\end{cases}
\end{align*}
For all non-negative integers $k\geq r$, the number of rank $r$ quadratic forms on an $\Fq$-vector space of dimension $k$ is
$$
N(k,r)=\gbinom{k}{r}N(r),
$$
where $\gbinom{k}{r}$
denotes the $q$-ary Gaussian binomial coefficient (see Definition~\ref{def:gaussian}).
\end{Theorem}

A more general result implying Theorem~\ref{thm:cor:counting-q:main} appears in~\cite[Chapter 6, Section 2]{Lidl}.

As to Theorem~\ref{thm:givenrank-gen:main}, the following references need to be mentioned.
In \cite[Lemma 9.5.9]{Brouwer} the number of symmetric bilinear forms of given rank is computed. In the odd characteristic case, as symmetric bilinear forms correspond to quadratic forms and the two notions of rank coincide, this result is equivalent to Theorem~\ref{thm:givenrank-gen:main}. 
As to the arbitrary characteristic case, \cite{Brouwer} refers to \cite{Egawa}. The latter uses the language of association schemes and gives a result that allows to compute (even though this is not explicitly stated) the number $N'(k,s)$ of quadratic forms of rank $r\in\{2s-1,2s\}$ on an $\Fq$-vector space of dimension $k$. This result is slightly weaker than our theorem, as it allows to compute the sum $N(k,2s-1)+N(k,2s)$ instead of $N(k,2s-1)$ and $N(k,2s)$ separately, but it would be sufficient for the main purpose of this work.

\section{Proof of Main Theorem~\ref{mthm:main}}\label{sec:proof}

We recall the notation introduced in Section~\ref{sec:statement}.
Given an $[n,k]$-code $C$ and denoting by $\pi_1,\dots,\pi_n\in\Fq^k$ the {\em columns} of a generator matrix of $C$ (i.e.\ a matrix whose {\em rows} form a basis of $C$), we define the linear map
$$
\begin{matrix}
\ev_C\colon&\Quad(\Fq^k)&\to&\F_q^n,\\
&Q&\mapsto&(Q(\pi_1),\dots,Q(\pi_n))
\end{matrix}
$$
whose image is $C^{*2}$.

Recall that we have defined the random variable $X(n,k):=|\ker\ev_C|$, with
distribution induced by a uniform random selection of $C$ from
$\CC(n,k)$.
For simplicity, we will write $X_k$ as a shorthand for $X(k(k+1)/2,k)$.

It is convenient to measure ``how far'' $C^{*2}$ is from being the
full space by defining, for all positive integers $n\geq k$ and all
non-negative integers $\ell$,
the probabilities:
$$
p_{\ell}(n,k):=\Pr(\codim C^{*2}\leq \ell),
$$
where $C$ is chosen uniformly at random from $\CC(n,k)$.
Using this notation, Main Theorem~\ref{mthm:main} claims that there exists $\delta\in\R_{>0}$ such that, for all large enough $k$, $p_0(n(k),k)\geq1-2^{-\delta t(k)}$.

As mentioned before, crucial to
the proof of Main Theorem~\ref{mthm:main} is to
estimate the expected value of $X_k=X(k(k+1)/2,k)$: 
this is precisely the purpose of Theorem~\ref{thm:expectation}, that states that
$\lim_{k\to\infty}\E\left[X_k\right]=2.$ We now proceed to its proof.

\medskip

\begin{Proof}[ of Theorem~\ref{thm:expectation}]
In Section~\ref{sec:statement} we defined the space $S$ of all quadratic forms vanishing at all unit vectors and we proved that, for all positive integers $m\geq k$,
\begin{equation*}\tag{\ref{eq0}}
\E[X(m,k)]=\sum_{Q\in S}{\left(\frac{|Z(Q)|}{q^k}\right)}^{m-k}.
\end{equation*}

We now fix a rank threshold, i.e.\ a fraction of $k$, and we classify the forms in $S$ accordingly. Precisely, for any $0<\alpha<1$ we define
\begin{gather*}
S^-(\alpha):=\{Q\in S:0<\rk Q\leq\alpha k\},\\ S^+(\alpha):=\{Q\in S:\rk Q>\alpha k\},
\end{gather*}
so $S=\{0\}\cup S^+(\alpha)\cup S^-(\alpha)$.
We observe that
\begin{equation}\label{eq1.1}
|S^-(\alpha)|
\leq q^{(-\frac{\alpha^2}{2}+\alpha)k^2+o(k^2)}.
\end{equation}
Indeed, by Theorem~\ref{thm:givenrank-gen:main} we have $$|S^-(\alpha)|=\sum_{r=1}^{\alpha k}N(k,r)=\sum_{r=1}^{\alpha k}\gbinom{k}{r}N(r).$$
We loosely bound $\gbinom{k}{r}\leq q^{r(k-r+1)}$ and $N(r)\leq|\Quad(\Fq^r)|=q^{r(r+1)/2}$ and we obtain
\begin{align*}
|S^-(\alpha)|&\leq\sum_{r=1}^{\alpha k}q^{r(k-r+1)}q^{r(r+1)/2}=
\sum_{r=1}^{\alpha k}q^{-\frac{r^2}{2}+(k+\frac32)r}\leq\\
&\leq\alpha k q^{(-\frac{\alpha^2}{2}+\alpha)k^2+\frac32\alpha k},
\end{align*}
proving~\eqref{eq1.1}.
This yields
$$
\frac{|S^-(\alpha)|}{|S|}\leq\frac{q^{(-\frac{\alpha^2}{2}+\alpha)k^2+o(k^2)}}{q^{\frac{k(k-1)}{2}}}
=q^{-\frac12{(\alpha-1)}^2k^2+o(k^2)}
$$
which tends to $0$ as $k\to\infty$. Hence, noting that $|S^+(\alpha)|=|S|-1-|S^-(\alpha)|$, we obtain
\begin{equation}\label{eq1.2}
\lim_{k\to\infty}\frac{|S^+(\alpha)|}{|S|}=1.
\end{equation}
In view to using the observations \eqref{eq1.1} and \eqref{eq1.2} 
on the ``density'' of $S^+(\alpha)$ and $S^-(\alpha)$ in $S$, we apply
the partition of $S$ to \eqref{eq0} and write
\begin{align}
\E&[X(m,k)]=\nonumber\\
&=1+\sum_{Q\in S^+(\alpha)}{\left(\frac{|Z(Q)|}{q^k}\right)}^{m-k}+\sum_{Q\in S^-(\alpha)}{\left(\frac{|Z(Q)|}{q^k}\right)}^{m-k}.\label{eq2}
\end{align}
We now prove that the first sum tends to $1$ while the second one (for some suitable value of $\alpha$) tends to $0$.

By Theorem~\ref{thm:cor:counting-q:main}, the number of zeros of any form $Q\in S^+(\alpha)$ is bounded by
$$
|Z(Q)|\leq q^{k-1}+(q-1)q^{k-\frac{\alpha k}{2}-1}\leq q^{k-1}\left(1+\frac{1}{q^{\frac{\alpha k}{2}-1}}\right)
$$
and
$$
|Z(Q)|\geq q^{k-1}-(q-1)q^{k-\frac{\alpha k}{2}-1}\geq q^{k-1}\left(1-\frac{1}{q^{\frac{\alpha k}{2}-1}}\right).
$$
It follows that
$$
\frac{1}{q}\left(1-\frac{1}{q^{\frac{\alpha k}{2}-1}}\right)\leq\frac{|Z(Q)|}{q^k}\leq\frac{1}{q}\left(1+\frac{1}{q^{\frac{\alpha k}{2}-1}}\right)
$$
hence
\begin{align*}
{\left(1-\frac{1}{q^{\frac{\alpha k}{2}-1}}\right)}^{m-k}\frac{|S^+(\alpha)|}{q^{m-k}}&\leq\sum_{Q\in S^+(\alpha)}{\left(\frac{|Z(Q)|}{q^k}\right)}^{m-k}\leq\\
&\leq{\left(1+\frac{1}{q^{\frac{\alpha k}{2}-1}}\right)}^{m-k}\frac{|S^+(\alpha)|}{q^{m-k}}.
\end{align*}
Setting $m=k(k+1)/2$, we get
\begin{align*}
{\left(1-\frac{1}{q^{\frac{\alpha k}{2}-1}}\right)}^\frac{k(k-1)}{2}&\frac{|S^+(\alpha)|}{|S|}\leq\sum_{Q\in S^+(\alpha)}{\left(\frac{|Z(Q)|}{q^k}\right)}^\frac{k(k-1)}{2}\leq\\
&\leq{\left(1+\frac{1}{q^{\frac{\alpha k}{2}-1}}\right)}^\frac{k(k-1)}{2}\frac{|S^+(\alpha)|}{|S|}.
\end{align*}
So the first sum in \eqref{eq2} is bounded, from above and from below, by functions which tend to $1$ (by~\eqref{eq1.2}), hence it tends to $1$, too.

We now prove that if we take any $0<\alpha<1-\sqrt{\log_q(2q-1)-1}$,
the last sum in \eqref{eq2} tends to $0$, which will conclude the
proof of the theorem.

By Theorem~\ref{thm:cor:counting-q:main},
all forms $Q\in S^-(\alpha)$ satisfy
$$
|Z(Q)|\leq q^{k-1}+(q-1)q^{k-2}=2q^{k-1}-q^{k-2}.
$$
This is trivial for odd rank forms, as they always have exactly $q^{k-1}$ zeros. We get
$$
\sum_{Q\in S^-(\alpha)}{\left(\frac{|Z(Q)|}{q^k}\right)}^{m-k}
\leq{\left(\frac{2q-1}{q^2}\right)}^{m-k}|S^-(\alpha)|.
$$
Setting $m=k(k+1)/2$ and using \eqref{eq1.1} we finally obtain
\begin{align*}
\sum_{Q\in S^-(\alpha)}&{\left(\frac{|Z(Q)|}{q^k}\right)}^{m-k}\leq\\
\leq&{\left(\frac{2q-1}{q^2}\right)}^{\frac{k(k-1)}{2}}q^{(-\frac{\alpha^2}{2}+\alpha)k^2+o(k^2)}=
q^{\mu(\alpha)k^2+o(k^2)},
\end{align*}
where $\mu(\alpha):=-\frac12(\alpha^2-2\alpha+2-\log_q(2q-1))<0$ under the assumptions on $\alpha$. Therefore the right
hand side tends to $0$. This concludes the proof.
\end{Proof}

As a first consequence of Theorem~\ref{thm:expectation}, we derive 
a proof of Theorem~\ref{thm:biglength}.

\begin{Proof}[ of Theorem~\ref{thm:biglength}]
As before, set $m(k):=k(k+1)/2$. Given a code $C\in\CC(n(k),k)$, we obtain a code $C'\in\CC(m(k),k)$ puncturing the last $s(k)$ coordinates of $C$.
We define $\NN$ to be the event ``$\dim C^{*2}=m(k)$'' and, for all $j\in\N$, we define $\EE_j$ to be the event ``$|\ker\ev_{C'}|=j$''.
We observe that $\dim C^{*2}=m(k)$ if and only if $\ker\ev_C=0$, and this holds if and only if for all nonzero $Q\in\ker\ev_{C'}$ there exists $i\in\{m(k)+1,\dots,n(k)\}$ such that $Q(\pi_i)\not=0$.
Hence, if in the case of $\EE_j$ we write $\ker\ev_{C'}\setminus\{0\}=\{Q_1,\dots,Q_{j-1}\}$, we have
\begin{align*}
\Pr(&\overline{\NN}|\EE_j)=\\
&=\Pr\left(\bigcup_{i=1}^{j-1}\left\{Q_i(\pi_{m(k)+1})=\dots=Q_i(\pi_{n(k)})=0\right\}\right)\leq\\
&\leq\sum_{i=1}^{j-1}{\Pr(Q_i(\pi)=0)}^{s(k)},
\end{align*}
for all $j\in\N$, where $\pi\in\Fq^k$ is chosen uniformly at random.
Moreover, for any nonzero quadratic form $Q\in\Quad(\Fq^k)$,
$$
\Pr(Q(\pi)=0)\leq\frac{q^{k-1}+(q-1)q^{k-2}}{q^k}=\frac{2q-1}{q^2}.
$$
Note that $(2q-1)/q^2$ is a constant strictly smaller than $1$.
It follows that
$$
\Pr(\overline{\NN}|\EE_j)\leq\sum_{i=1}^{j-1}{\left(\frac{2q-1}{q^2}\right)}^{s(k)}=(j-1){\left(\frac{2q-1}{q^2}\right)}^{s(k)}.
$$
Applying the law of total probability to $\Pr(\overline{\NN})$ together with the above observations we finally have
\begin{align*}
\Pr(\overline{\NN})&=\sum_{j\in\N}\Pr(\EE_j)\Pr(\overline{\NN}|\EE_j)\leq\\
&\leq{\left(\frac{2q-1}{q^2}\right)}^{s(k)}\sum_{j\in\N}\Pr(\EE_j)(j-1)=\\
&={\left(\frac{2q-1}{q^2}\right)}^{s(k)}(\E[X_k]-1).
\end{align*}
The conclusion follows by Theorem~\ref{thm:expectation}.
\end{Proof}

Next, we derive from the estimation of the expectation of $X_k$ given
by Theorem~\ref{thm:expectation}, 
a lower bound for the probability of $X_k$ being smaller than some fixed constant. Precisely, the following holds.

\begin{Proposition}\label{prop:probgen}
For any $\varepsilon>0$ there exists $k_\varepsilon\in\N$ such that, for all $k\geq k_\varepsilon$, for every non-negative integer $\ell$ we have
$$
\Pr\left(\dim C^{*2}\geq\frac{k(k+1)}{2}-\ell\right)\geq1-\frac{2+\varepsilon}{q^{\ell+1}},
$$
where $C$ is chosen uniformly at random from $\CC(k(k+1)/2,k)$.
\end{Proposition}

\begin{Proof}
We apply Markov's inequality to the random variable $X_k$, namely:
\begin{equation}
  \label{eq:Markov}
  \Pr(X_k<\delta)\geq1-\frac{\E[X_k]}{\delta}
\end{equation}
for any $\delta>0$.
By Theorem~\ref{thm:expectation} there exists $k_\varepsilon\in\N$
such that, for all $k\geq k_\varepsilon$, we have
$\E\left[X_k\right]\leq2+\varepsilon$, hence for any $\delta>0$,
\eqref{eq:Markov} gives
$$
\Pr(X_k<\delta)\geq1-\frac{2+\varepsilon}{\delta}
$$
if $k\geq k_\varepsilon$. Now setting $\delta=q^{\ell+1}$ and noting that $\Pr(X_k<q^{\ell+1})=\Pr(\dim C^{*2}\geq k(k+1)/2-\ell)$ we conclude.
\end{Proof}

Proposition~\ref{prop:probgen} together with
Proposition~\ref{prop:distdualsq} allow us to conclude 
the proof of Main Theorem~\ref{mthm:main}.

\medskip

\begin{Proof}[ of Main Theorem~\ref{mthm:main}]
%
Let $k\leq n< m:=k(k+1)/2$ be positive integers, and let $t:=m-n$. We
use a puncturing argument.
The key observation is that a random code of length $n$ can be
obtained by first choosing a random code of length $m$ and then
deleting $m-n$ random coordinates. We shall look closely at the
probability that non-zero words survive in the dual of the punctured code.

Precisely, consider a uniform random code $C\in\CC(m,k)$: let 
$C'\in\CC(n,k)$ be obtained from $C$ by removing $t$ random coordinates among
the last $m-k$. Let these $t$ coordinates be chosen uniformly,
independently of $C$.

In order to estimate $p_0(n,k)$, we define the following events.
Call $\EE$ the event studied in Proposition~\ref{prop:distdualsq},
namely $d_{\min}((C^{*2})^\bot)\leq cm$ where $c$ is the constant of Proposition~\ref{prop:distdualsq}.
For all non-negative integers $i$, call $\EE_i$ the event $\codim C^{*2}=i$. 
As before, bar denotes the complement event.

For any positive integer $\ell$ we have
\begin{equation}
  \label{eq:EE}
  p_0(n,k)\geq\sum_{i=1}^\ell\Pr(\overline\EE\cap\EE_i)
\Pr(\codim (C')^{*2}=0|\overline\EE\cap\EE_i).
\end{equation}
Let $C_0$ be a fixed code of length $m$ and suppose $x$ is a codeword of
$C_0^{\perp}$ of weight $w$. Puncture $C_0$ by removing $t$ random
coordinates among the last $m-k$.
The probability that none of the random $t$ coordinates belong to the support of $x$ is at most
\begin{equation}
  \label{eq:binomial}
  \frac{\binom{m-w}{t}}{\binom{m-k}{t}}
\end{equation}
(and actually equal to~\eqref{eq:binomial} if the support of $x$ contains the first $k$ coordinates). 
If the dual code
$C_0^{\perp}$ contains exactly  $q^i-1$ non-zero codewords all of which
have weight at least $cm$, then the probability that the $t$ random
coordinates miss the support of at least one codeword of $C_0^{\perp}$
is, by \eqref{eq:binomial} and the union bound, bounded from above by
$$(q^i-1)\frac{\binom{m-cm}{t}}{\binom{m-k}{t}}.$$
Now observing that a non-zero codeword in
$((C')^{*2})^\bot$ exists only if there exists a non-zero codeword in
$(C^{*2})^\bot$ with support disjoint from the chosen $t$
coordinates, we obtain that, for all $i=1,\dots,\ell$,
\begin{align*}
\Pr(\codim (C')^{*2}\not=0|\overline\EE\cap\EE_i)&\leq 
(q^i-1)\frac{\binom{m-cm}{t}}{\binom{m-k}{t}}\leq\\
&\leq q^\ell\ \frac{\binom{m-cm}{t}}{\binom{m-k}{t}}.
\end{align*}
We bound the fraction as follows:
\begin{align*}
\frac{\binom{m-cm}{t}}{\binom{m-k}{t}}&=\frac{(m-cm)\cdots(m-cm-t+1)}{(m-k)\cdots(m-k-t+1)}\leq\\
&\leq{\left(\frac{m-cm}{m-k}\right)}^t={(1-c)}^t{\left(\frac{k+1}{k-1}\right)}^t
\end{align*}
from which we obtain
$$
\Pr(\codim (C')^{*2}\not=0|\overline\EE\cap\EE_i)\leq q^{\ell +t(\log(1-c)+\log \frac{k+1}{k-1})}.
$$
Since $\log \frac{k+1}{k-1}$ goes to zero when $k$ goes to infinity
and $\log(1-c)$ is negative, by fixing $\ell=\alpha t$
we get the existence of a positive $\beta$ such that, for any $k$
large enough,
\begin{equation}
  \label{eq:betat}
  \Pr(\codim (C')^{*2}\not=0|\overline\EE\cap\EE_i)\leq q^{-\beta t}.
\end{equation}
Now note that by the union bound
\begin{align*}
\Pr(\overline\EE\cap\EE_i)&=1-\Pr(\EE\cup\overline\EE_i)
\geq1-\Pr(\EE)-\Pr(\overline\EE_i)=\\&=\Pr(\EE_i)-\Pr(\EE).
\end{align*}
Therefore, \eqref{eq:betat} with \eqref{eq:EE} give
\begin{align}
  p_0(n,k&)\geq (1-q^{-\beta
    t})\sum_{i=1}^{\ell}(\Pr(\EE_i)-\Pr(\EE))\nonumber\\
          &\geq (1-q^{-\beta t})(1-\Pr(\dim C^{*2}\leq m-\ell) - \ell\Pr(\EE)).\label{eq:ellEE}
\end{align}
Proposition~\ref{prop:probgen} gives us, since $\ell=\alpha t$,
that $\Pr(\dim C^{*2}\leq m-\ell)\leq 2^{\beta 't}$ for a constant
$\beta'$.
Proposition~\ref{prop:distdualsq} gives us, since $\ell\leq k^2$, that
$\ell\Pr(\EE)\leq 2^{-\gamma k}$ for some constant $\gamma$.
From~\eqref{eq:ellEE} we therefore get
$$
p_0(n,k)\geq1-2^{-\gamma k}-2^{-\delta t}.
$$
for constants $\gamma$ and $\delta$.
\end{Proof}

\section{Changing the probabilistic model}\label{sec:prob_model}

In this section we expand Remark~\ref{rem:prob_model},
with the purpose of showing that, even though our probabilistic model may appear restrictive, our analysis gives all the ingredients necessary to consider different models.

For all positive integers $n\geq k$ we define the following two families of codes.
Let $\AAA(n,k)$ be the family of all codes of length $n$ and dimension at most $k$ with the following distribution: choose a $k\times n$ matrix $A$ uniformly at random and pick the code spanned by the rows of $A$. 
Let $\UU(n,k)$ be the family of all codes of length $n$ and dimension
$k$, with uniform distribution. Note that it is equivalent to a uniform random choice of a $k\times n$ full-rank matrix, as each such a code has the same number of bases, hence the same number of generator matrices.

We first argue that all our results hold if we replace $\CC(n,k)$ with
$\AAA(n,k)$. The two probability distributions are subtly different
and it is not easy to derive results for $\AAA(n,k)$ from the results
for $\CC(n,k)$ seen as ``black boxes''. However, if we go over the
proofs of our theorems, we see that they will carry over to
$\AAA(n,k)$ with no significant change of strategy. Specifically, in
the proof of Theorem~\ref{thm:expectation}, one will replace the study
of the quantity $\sum_{Q\in S}{\left(\frac{|Z(Q)|}{q^k}\right)}^{m-k}$
in \eqref{eq0} by 
$$\sum_{Q}{\left(\frac{|Z(Q)|}{q^k}\right)}^{m}$$
where $Q$ ranges over all quadratic forms on $k$ variables. The
quantity to be studied is simply the expected number of quadratic
forms that vanish on $m$ random values. Going over the proof one will
end up with exactly the same expected value. We sum over a space with $q^k$
more quadratic forms but replace probabilities of the form
$(|Z(Q)|/q^k)^{m-k}$ by $(|Z(Q)|/q^k)^m$ which behaves like $1/q^k$
times less.
Regarding the probabilistic analysis that proves
Proposition~\ref{prop:distdualsq},
we see that it is virtually unchanged when the first $k$ coordinates
become random. Also the puncturing argument that proves
Theorem~\ref{mthm:main} sees only the punctured coordinates being
chosen from $\{1,\ldots ,m\}$ rather than from $\{k+1,\ldots ,m\}$.

Regarding the second distribution $\UU(n,k)$, we argue differently and
relate it to $\AAA(n,k)$.
From here on $n$ and $k$ will be suppressed from the notation,
since they are assumed to be fixed. 
We add indices as $C\leftarrow\AAA$ or $C\leftarrow\UU$ to our probability notation to make the probabilistic model explicit.
Observe that for any fixed code $C_0$ of dimension $k$, we have
$$
\Pr_{C\leftarrow\AAA}(C=C_0|\dim C=k) = \Pr_{C\leftarrow\UU}(C=C_0).
$$
It follows that, if $\PP(C)$ denotes a property that a code $C$ may have,
$$
\Pr_{D\leftarrow\UU}(\PP(D))=\Pr_{C\leftarrow\AAA}(\PP(C)|\dim C=k).
$$
We deduce from this observation that:
\begin{Lemma}\label{lem:AandU}
For any property $\PP$,
$$
\Pr_{D\leftarrow\UU}(\PP(D))\geq \Pr_{C\leftarrow \AAA}(\PP(C))-\Pr_{C\leftarrow\AAA}(\dim C<k).
$$
\end{Lemma}
\begin{Proof}
We have
\begin{align*}
\Pr_{C\leftarrow \AAA}(\PP(C))&=
\Pr_{C\leftarrow \AAA}(\PP(C)|\dim C=k)\Pr_{C\leftarrow \AAA}(\dim C=k)+\\
&+\Pr_{C\leftarrow \AAA}(\PP(C)|\dim C<k)\Pr_{C\leftarrow \AAA}(\dim C<k)\leq\\
&\leq\Pr_{D\leftarrow\UU}(\PP(D))+\Pr_{C\leftarrow \AAA}(\dim C<k).
\end{align*}
\end{Proof}

Next, recall this well-known result on random matrices:
$$
\Pr_{C\leftarrow\AAA}(\dim C<k)\leq \frac{1}{q^{n-k}}.
$$
Together with Lemma~\ref{lem:AandU} this gives us:
$$\Pr_{D\leftarrow\UU}(\PP(D))\geq \Pr_{C\leftarrow \AAA}(\PP(C)) -
\frac{1}{q^{n-k}}.$$ 
We can now apply this to versions of our Theorems for $\AAA(n,k)$.
In particular, our main Theorem~\ref{mthm:main} will read, under the
uniform distribution $\UU(n,k)$, that there exist  some positive real
constants $\gamma,\delta$ such that
$$
\Pr_{C\leftarrow\UU}(C^{*2}=\Fq^{n(k)})\geq1-2^{-\gamma k}-2^{-\delta t(k)}-\frac{1}{q^{n(k)-k}}.
$$
This simple argument is enough to recover an asymptotically optimal
version of our main result for the uniform distribution, except for code rates that tend to $1$.

\appendix

\section{Quadratic forms}\label{sec:appquadforms}

This appendix is meant to be a continuation of Section~\ref{sec:quadforms}.
In particular, we refer to that section for the definitions of quadratic form, radical and rank.

{\em Let $K$ be a field, let $(V,Q)$ be a $K$-quadratic space.}
%

With abuse of terminology, $V$ itself is called a quadratic space.
Recall that $V$, as a vector space, is finite dimensional by definition.
Any subspace $W$ of $V$ inherits a natural structure of quadratic space, defined by the restriction of $Q$ to $W$.

Recall that we defined a symmetric bilinear form $\tilde B_Q$ on $V$ by
$$
\tilde B_Q(x,y):=Q(x+y)-Q(x)-Q(y)
$$
for all $x,y\in V$. 
If $\charf K\not=2$ we also define the symmetric bilinear form $B_Q:=\frac{1}{2}\tilde B_Q$, which satisfies $B_Q(x,x)=Q(x)$ for all $x\in V$. If $\charf K=2$ note that $\tilde B_Q$ is alternating, i.e.\ $\tilde B_Q(x,x)=0$ for all $x\in V$.
As a shorthand, if there is no ambiguity we write $x\cdot y$ instead of $\tilde B_Q(x,y)$ for $x,y\in V$.

A remark concerning the definitions of radical and rank follows.
%
If $\charf K\not=2$ then $Q$ vanishes on $\Rad V$: indeed, for all $x\in\Rad V$ we have $Q(x)=B_Q(x,x)=\frac12x\cdot x=0$ by definition of the radical.
If $\charf K=2$ this is not always the case: for example, consider the quadratic form on $\F_2$ defined by $Q(x):=x^2$; 
note that $\widetilde B_Q$ is identically zero, hence the radical is the whole space, but $Q$ does not vanish at $x=1$.
So in the characteristic $2$ case $\Rad^0V$, the zero locus of the restriction of $Q$ to $\Rad V$, is not necessarily trivial.
Following \cite{Dieudonne}, we have defined the rank of a quadratic form to be the codimension of this zero locus.
%

In the characteristic $2$ case, under the additional assumption that $K$ is perfect, i.e.\ squaring is an automorphism of $K$ (which is always the case if $K$ is a finite field), one can prove that the difference between the rank of $Q$ and the codimension of the radical of $V$
is either zero or one.


We define orthogonality and isotropy with respect to $\tilde B_Q$, as follows.

Two vectors $x,y\in V$ are orthogonal if $x\cdot y=0$. Two subspaces $V_1,V_2\subseteq V$ are orthogonal if $x\cdot y=0$ for all $x\in V_1,y\in V_2$.
We use the symbol $\perp$ for the orthogonality relation.
The orthogonal of a subspace $V_1\subseteq V$ is
$$
V_1^\perp:=\{x\in V:x\cdot y=0\text{ for all }y\in V_1\}.
$$
Note that $V_1\cap V_1^\perp=\Rad V_1$, so $\Rad V_1=0$ implies $V_1\cap V_1^\perp=0$. Moreover, by basic linear algebra $\dim V_1+\dim V_1^\perp=\dim V$. Hence in this case $V_1^\perp$ is a complement of $V_1$, called the orthogonal complement of $V_1$.
Finally, a decomposition of $V$ is orthogonal if the components are pairwise orthogonal.

A non-zero vector $x\in V$ is isotropic if $x\cdot x=0$. A subspace of $V$ is isotropic if it contains an isotropic vector, anisotropic otherwise. Note that if $\charf K=2$ then every vector is isotropic, as $\tilde B_Q$ is alternating, hence it does not make sense to use this notion.

A quadratic space $(V,Q)$ is classified according to the orthogonal decomposition induced on $V$ by $Q$. The ``building blocks'' in this decomposition are hyperbolic and symplectic planes, that are defined below.

\begin{Definition}[hyperbolic plane]
Assume that $\charf K\not=2$.
A hyperbolic plane is a non-degenerate $2$-dimensional subspace which admits a basis of isotropic vectors.
\end{Definition}

Note that any hyperbolic plane $H$ admits a basis $\{v_1,v_2\}$ of isotropic vectors such that $v_1\cdot v_2=1$. Indeed, for any basis $\{v_1,w\}$, with $v_1,w$ isotropic, it holds that $\alpha:=v_1\cdot w\not=0$ as $H$ is non-degenerate, hence $\{v_1,v_2\}$ with $v_2:=\alpha^{-1}w$ satisfies the property.


\begin{Theorem}[Witt's decomposition]\label{thm:class-odd}
Assume that $\charf K\not=2$.
Then the quadratic space $V$ orthogonally decomposes as
$$
V=\Rad V\oplus \bigoplus_{i=1}^m H_i\oplus W,
$$
where the $H_i$'s are hyperbolic planes and $W$ is anisotropic. Moreover, if $K$ is finite then $\dim W\leq 2$.
\end{Theorem}
\begin{Proof}
Any complement of $\Rad V$ is non-degenerate and orthogonal to $\Rad V$, so we may assume that $\Rad V=0$, i.e.\ $V$ is non-degenerate. If $V$ is anisotropic we are done, with $m=0$ and $V=W$. Otherwise there exists an isotropic vector $v_1\in V$, hence $x\in V$ such that $\alpha:=v_1\cdot x\not=0$, as $V$ is non-degenerate. Now take
$$
v_2:=\frac{1}{\alpha}x-\frac{x\cdot x}{2\alpha^2}v_1,
$$
$H_1:=\langle v_1,v_2\rangle$ and apply induction.

If $K$ is finite then $\dim W\leq 2$, as any quadratic form on a non-degenerate space of dimension larger than $2$ has a non trivial zero, which is an isotropic vector of $V$. This is a consequence of the Chevalley-Warning Theorem, see for example \cite{Serre}.
\end{Proof}

\begin{Definition}[symplectic plane]
Assume that $\charf K=2$.
A symplectic plane is a subspace which admits a basis $\{v_1,v_2\}$ such that $v_1\cdot v_2=1$.
\end{Definition}

Non-degeneracy is implied by this definition.

\begin{Theorem}\label{thm:class-2}
Assume that $\charf K=2$.
Then the quadratic space $V$ orthogonally decomposes as
$$
V=\Rad V\oplus \bigoplus_{i=1}^m S_i,
$$
where the $S_i$'s are symplectic planes. Moreover, all but at most one among the $S_i$'s admit a $K$-basis $\{v_1,v_2\}$ such that $v_1\cdot v_2=1$ and $Q(v_1)=Q(v_2)=0$.
\end{Theorem}
\begin{Proof}
Again, we may assume that $V$ is non-degenerate. Let $v_1\in V$, let $x\in V$ be such that $\alpha:=v_1\cdot x\not=0$. Take $v_2:=\frac{1}{\alpha}x$, $S_1:=\langle v_1,v_2\rangle$ and argue by induction.
For the last statement, see \cite{Dye} or \cite[Chapter I, Section 16]{Dieudonne}.
\end{Proof}


\begin{Remark}
Stronger results actually hold. The decompositions above are, in some sense, unique: for example, in a Witt decomposition, the number $m$ of hyperbolic planes is unique while the anisotropic space $W$ is unique up to ``isometry''. For details, see \cite{Lam,Serre} for Theorem~\ref{thm:class-odd} and \cite{Dye,Dieudonne} for Theorem~\ref{thm:class-2}. However, these stronger results are not needed here.
\end{Remark}

\subsection{Number of zeros of a quadratic form}

{\em Let $(V,Q)$ be a quadratic space over the finite field $\Fq$.}

In this section we compute the number of zeros in $V$ of the quadratic form $Q$, as a function of the dimension $k$ of $V$, the rank $r$ of $Q$ and the cardinality $q$ of the base field. Even though the definition of rank is essentially dependent on $\charf\Fq$, the formula we give is characteristic-free.

\begin{Theorem}\label{thm:cor:counting-q}
The number of vectors $x\in V$ such that $Q(x)=0$ is
\begin{enumerate}[a.]
\item $q^{k-1}$ if $r$ is odd,
\item either $q^{k-1}-(q-1)q^{k-\frac{r}{2}-1}$ or $q^{k-1}+(q-1)q^{k-\frac{r}{2}-1}$ if $r$ is even.
\end{enumerate}
\end{Theorem}

\begin{Remark}
The ``$\pm$'' in claim b of Theorem~\ref{thm:cor:counting-q} (and of the forthcoming Theorem~\ref{thm:counting-q}) only depends on the ``last component'' in the orthogonal decomposition of $V$ given by Theorem~\ref{thm:class-odd} or Theorem~\ref{thm:class-2}.
\end{Remark}

In \cite[Chapter 6, Section 2]{Lidl} the number of vectors $x\in V$ such that $Q(x)=b$, for any full-rank quadratic form $Q$ on $V$ and any $b\in\Fq$, is computed. Theorem~\ref{thm:counting-q} below, whence Theorem~\ref{thm:cor:counting-q} easily follows, is an instance of this result.
However, for completeness, and to show an application of the classification theorems, we include a full proof of Theorem~\ref{thm:counting-q}.


Here, it is convenient to view quadratic forms as polynomials, as follows. This correspondence holds over an arbitrary field $K$ (so we abandon for a moment the assumption that the base field is finite).
Fixing a $K$-basis $\{v_1,\dots,v_k\}$ of $V$ we can associate to $Q$ a homogeneous quadratic $k$-variate polynomial $f_Q\in K[X_1,\dots,X_k]$ such that, for all $(\alpha_1,\dots,\alpha_k)\in K^k$,
$$
Q(\alpha_1v_1+\dots+\alpha_kv_k)=f_Q(\alpha_1,\dots,\alpha_k),
$$
namely
$$
f_Q:=\sum_{1\leq i\leq k}Q(v_i)X_i^2+\sum_{1\leq i<j\leq k}\tilde B_Q(v_i,v_j)X_iX_j.
$$
Clearly there is a one-to-one correspondence between zeros of $Q$ and zeros of $f_Q$, independently of the basis choice.
We remark that the rank of $Q$ can be equivalently defined as the minimal number of variables appearing in the polynomial $f_Q$ associated to $Q$, where minimality is taken over all possible basis choices.

Back to the case of $K=\Fq$, we have the following straightforward consequence of the classification theorems.

\begin{Corollary}\label{cor:class}
Assume that $r\geq3$. Then the polynomial $f_Q$ associated to $Q$ in some suitable basis can be written as
$$
f_Q=g_Q+X_{k-1}X_k,\quad\quad\text{with}\quad\quad g_Q\in\Fq[X_1,\dots,X_{k-2}].
$$
\end{Corollary}
\begin{Proof}
As $r\geq3$, the classification theorems give an $\Fq$-basis $\{v_1,\dots,v_k\}$ of $V$ such that $\tilde B_Q(v_{k-1},v_k)=1$, $Q(v_{k-1})=Q(v_k)=0$ and $\langle v_1,\dots,v_{k-2}\rangle\perp\langle v_{k-1},v_k\rangle$.
The polynomial $f_Q$ associated to $Q$ with respect to this basis has the desired form.
\end{Proof}

We are ready to proceed.
We start with the case of full-rank forms, and then we show how the general case easily follows. 

\begin{Theorem}\label{thm:counting-q}
Assume that $r=k$, i.e.\ that $Q$ has full rank.
Then the number of vectors $x\in V$ such that $Q(x)=0$ is
\begin{enumerate}[a.]
\item $q^{k-1}$ if $k$ is odd,
\item either $q^{k-1}-(q-1)q^{\frac{k}{2}-1}$ or $q^{k-1}+(q-1)q^{\frac{k}{2}-1}$ if $k$ is even.
\end{enumerate}
\end{Theorem}
\begin{Proof}
Denote by $Z_k(f)$ the number of zeros in $\Fq^k$ of a polynomial $f\in\Fq[X_1,\dots,X_k]$. The proof is by induction on $k$.
If $k=1$ (case {\em a}) then in some basis $f_Q=\alpha X_1^2$ and its only zero is the zero vector.
If $k=2$ (case {\em b}) then, by classification theorems, we have two possible situations: either the only zero of $f_Q$ is the zero vector or $f_Q=X_1X_2$ has $2q-1$ zeros.

Now let $k\geq3$. By Corollary~\ref{cor:class} we can write
$$
f_Q=g_Q+X_{k-1}X_k,\quad\quad\text{with}\quad\quad g_Q\in\Fq[X_1,\dots,X_{k-2}].
$$
Note that the zeros of $f_Q$ are exactly all $k$-tuples $(x,\alpha_1,\alpha_2)$ with $x\in\Fq^{k-2},\alpha_1,\alpha_2\in\Fq$ such that
\begin{itemize}
\item $x$ is a zero of $g_Q$ and $\alpha_1\alpha_2=0$ or
\item $x$ is not a zero of $g_Q$, $\alpha_1\not=0$ and $\alpha_2=-\alpha_1^{-1}g_Q(x)$.
\end{itemize}
Hence we get the recursion formula
\begin{align*}
Z_k(f_Q)&=(2q-1)Z_{k-2}(g_Q)+\\
&+(q-1)(q^{k-2}-Z_{k-2}(g_Q))=\\
&=q^{k-1}-q^{k-2}+qZ_{k-2}(g_Q)
\end{align*}
for $k\geq3$. This gives the result.
\end{Proof}

\begin{Proof}[ of Theorem~\ref{thm:cor:counting-q}]
In a suitable basis, the polynomial associated to $Q$ is $r$-variate, i.e.\ $f_Q\in\Fq[X_1,\dots,X_r]$.
This defines a full-rank quadratic form on $\Fq^r$, hence Theorem~\ref{thm:counting-q} applies.
The conclusion now follows as any zero of $f_Q$ in $\Fq^r$ gives $q^{k-r}$ zeros of $f_Q$ in $\Fq^k$ by padding.
\end{Proof}
%

\subsection{Number of quadratic forms of given rank}\label{sec:rank}

In this section we compute the number $N(k,r)$ of rank $r$ quadratic forms on any $\Fq$-vector space of dimension $k$, where $k,r$ are non-negative integers with $k\geq r$.
First we deal with the case $k=r$, i.e.\ of full-rank quadratic forms, then we address the general case. In the full-rank case we write $N(k)$ instead of $N(k,k)$, as a shorthand.
We now state the results: Theorem~\ref{thm:fullrank-gen} for the first case, Theorem~\ref{thm:givenrank-gen} for the latter.


\begin{Theorem}\label{thm:fullrank-gen}
For all non-negative integers $k$, the number of full-rank quadratic forms on an $\Fq$-vector space of dimension $k$ is
\begin{align*}
N(k)&=q^{\left\lfloor\frac{k}{2}\right\rfloor\left(\left\lfloor\frac{k}{2}\right\rfloor+1\right)}\prod_{i=1}^{\left\lceil\frac{k}{2}\right\rceil}(q^{2i-1}-1)=\\&=
\begin{cases}
q^{\frac{k-1}{2}\frac{k+1}{2}}\prod_{i=1}^\frac{k+1}{2}(q^{2i-1}-1)&\text{if }k\text{ is odd,}\\
q^{\frac{k}{2}\left(\frac{k}{2}+1\right)}\prod_{i=1}^\frac{k}{2}(q^{2i-1}-1)&\text{if }k\text{ is even.}
\end{cases}
\end{align*}
\end{Theorem}

\begin{Theorem}\label{thm:givenrank-gen}
For all non-negative integers $k\geq r$, the number of rank $r$ quadratic forms on an $\Fq$-vector space of dimension $k$ is
$$
N(k,r)=\gbinom{k}{r}N(r),
$$
where $\gbinom{k}{r}$
denotes the $q$-ary Gaussian binomial coefficient (see Definition~\ref{def:gaussian}).
\end{Theorem}

\begin{Remark}\label{rem:subsp}
Recall that $\gbinom{k}{r}$ equals the number of $r$-dimensional subspaces of any $\Fq$-vector space of dimension $k$. 
\end{Remark}


Our proofs of Theorems~\ref{thm:fullrank-gen} and~\ref{thm:givenrank-gen} follow.
Our strategy consists of constructing all quadratic forms on a given space as ``combinations'' (in the sense of Definition~\ref{def:quadraticformsum} and Construction~\ref{construction 1} below) of quadratic forms on subspaces. Counting recursively the number of forms constructed in this way and dividing by the number of repetitions will give the required quantity.

Towards a proof of Theorem~\ref{thm:fullrank-gen}, we fix a non-negative integer $k$ and an $\Fq$-vector space $V$ of dimension $k$. We define the following ``sum'' of quadratic forms.

\begin{Definition}\label{def:quadraticformsum}
Let $V_1,V_2\leq V$ be subspaces such that $V_1\cap V_2=0$, let $Q_1$ be a quadratic form on $V_1$ and $Q_2$ a quadratic form on $V_2$. We define $Q:=Q_1\oplus Q_2$ to be the unique quadratic form on $V_1\oplus V_2$ defined by the conditions ${Q|}_{V_1}=Q_1$, ${Q|}_{V_2}=Q_2$ and $V_1\perp V_2$.
\end{Definition}

In other words, for $v\in V_1\oplus V_2$, we define $Q(v):=Q_1(v_1)+Q_2(v_2)$, where $v_1\in V_1$ and $v_2\in V_2$ are the unique vectors such that $v_1+v_2=v$.
Also note that $\Rad(V_1\oplus V_2)=\Rad V_1\oplus\Rad V_2$.
So we construct quadratic forms on $V$ as follows.

\begin{Construction}\label{construction 1}
Let $h\leq k$ be a non-negative integer.
Let $(V_1,V_2,Q_1,Q_2)$ be a $4$-tuple consisting of a subspace $V_1\leq V$ of dimension $h$, a complement $V_2\leq V$ of $V_1$, a full-rank quadratic form $Q_1$ on $V_1$ and a full-rank quadratic form $Q_2$ on $V_2$. Define $Q:=Q_{(V_1,V_2,Q_1,Q_2)}:=Q_1\oplus Q_2\in\Quad(V)$.
\end{Construction}

The choice of the parameter $h$ is determined by the characteristic of $\Fq$ and the parity of the dimension $k$ of $V$, as follows:
\begin{enumerate}
\item $h=1$ if $k$ is odd and $\charf\Fq\not=2$,
\item $h=2$ if $k$ is even and $\charf\Fq\not=2$,
\item $h=2$ if $\charf\Fq=2$.
\end{enumerate}
We prove that, with this choice of $h$, all full-rank quadratic forms on $V$ are obtained by Construction~\ref{construction 1} and, conversely, all forms defined using Construction~\ref{construction 1} have full rank.

\begin{Lemma}\label{lemma:fullrank1}
Any full-rank quadratic form on $V$ is an instance of Construction~\ref{construction 1} with $h$ chosen as above.
\end{Lemma}
\begin{Proof}
First assume that $\charf\Fq\not=2$. If $Q$ is a full-rank quadratic form on $V$ then by Theorem~\ref{thm:class-odd} we have an orthogonal decomposition
$$
V=\bigoplus_{i=1}^mH_i\oplus W,
$$
with $\dim H_i=2$ for all $i=1,\dots,m$ and $\dim W\leq 2$.
If $k$ is odd then $\dim W$ is also odd, hence it must equal $1$. Let $V_1:=W$, $V_2:=\bigoplus_{i=1}^mH_i$, $Q_1:={Q|}_{V_1}$ and $Q_2:={Q|}_{V_2}$, then $Q=Q_{(V_1,V_2,Q_1,Q_2)}$ with $h=\dim W=1$.
If $k$ is even, let $V_1:=H_1,V_2:=\bigoplus_{i=2}^mH_i\oplus W,Q_1:={Q|}_{V_1},Q_2:={Q|}_{V_2}$, then $Q=Q_{(V_1,V_2,Q_1,Q_2)}$ with $h=\dim H_1=2$.

Now assume $\charf\Fq=2$. If $Q$ is a full-rank quadratic form on $V$ then by Theorem~\ref{thm:class-2} we have an orthogonal decomposition
$$
V=\Rad V\oplus\bigoplus_{i=1}^mS_i
$$
with $\dim\Rad V=0$ or $1$. Let $V_1:=S_1,V_2:=\Rad V\oplus\bigoplus_{i=2}^mS_i,Q_1:={Q|}_{V_1},Q_2:={Q|}_{V_2}$, then $Q=Q_{(V_1,V_2,Q_1,Q_2)}$ with $h=\dim S_1=2$.
\end{Proof}

\begin{Lemma}\label{lemma:fullrank2}
Any instance of Construction~\ref{construction 1}, with $h$ chosen as above, is a full-rank quadratic form on $V$.
\end{Lemma}
\begin{Proof}
Let $V_1,V_2,Q_1,Q_2$ be as in Construction~\ref{construction 1}, and let $Q:=Q_{(V_1,V_2,Q_1,Q_2)}$.
The statement is obvious if $\charf\Fq$ is odd: in this case both $\Rad V_1=\Rad V_2=0$, hence $\Rad(V_1\oplus V_2)=0$ as well.
The same happens in the characteristic $2$ case if both $h$ and $k$ are even.

The only non trivial case is the one of $\charf\Fq=2$ and $k$ odd. We have chosen $h$ to be even, hence $\Rad V_1=0$ while $\Rad V_2=\langle w\rangle$ for some $w\in V_2$ such that $Q(w)\not=0$.
Then $\Rad(V_1\oplus V_2)=\langle w\rangle$ and $Q(w)=Q_2(w)\not=0$, hence $Q$ has full rank.
\end{Proof}

It follows that the number of full-rank quadratic forms on $V$ is given by the number of suitable $4$-tuples $(V_1,V_2,Q_1,Q_2)$ divided by the number of repetitions.
The number of possible choices for $V_1$ is given by a Gaussian binomial coefficient. 
The following combinatorial lemma computes the number of possible choices for $V_2$.


\begin{Lemma}\label{lemma:0}
Let $h\leq k$ be a non-negative integer.
The number of complements of an $h$-dimensional subspace of $V$ is $q^{h(k-h)}$.
\end{Lemma}
\begin{Proof}
Let $W$ be an $h$-dimensional subspace of $V$, with basis $\{v_1,\dots,v_h\}$. This can be completed to a basis of $V$ in $(q^k-q^h)(q^k-q^{h+1})\cdots(q^k-q^{k-1})$ ways. Any complement of $W$ has dimension $k-h$, hence $(q^{k-h}-1)(q^{k-h}-q)\cdots(q^{k-h}-q^{k-h-1})$ different bases. Hence the number of complements of $W$ is
$$
\frac{q^k-q^h}{q^{k-h}-1}\cdot\frac{q^k-q^{h+1}}{q^{k-h}-q}\cdots\frac{q^k-q^{k-1}}{q^{k-h}-q^{k-h-1}}=
q^{h(k-h)}.
$$
\end{Proof}

Finally, we count how many times a quadratic form is repeated.

\begin{Lemma}\label{lemma:repetitions}
Let $Q$ be a full-rank quadratic form on $V$. For any non-degenerate $h$-dimensional subspace $V_1$ of $V$, with $h$ chosen as above, we have a unique complement $V_2$ of $V_1$ and unique full-rank quadratic forms $Q_1$ and $Q_2$ on $V_1$ and $V_2$ respectively such that $Q=Q_{(V_1,V_2,Q_1,Q_2)}$.
\end{Lemma}
\begin{Proof}
Let $V_1$ be a non-degenerate $h$-dimensional subspace of $V$. We want to define $V_2,Q_1,Q_2$ such that $Q_{(V_1,V_2,Q_1,Q_2)}=Q$. Clearly we have to take $Q_1:={Q|}_{V_1}$. The choice of $h$ implies that $\Rad V_1=0$, hence $V_1$ has an orthogonal complement. So take $V_2:=V_1^\perp$ and $Q_2:={Q|}_{V_2}$.
Note that these are the only possible choices, hence this proves the lemma.
\end{Proof}

For all full-rank quadratic forms $Q$ on $V$ and all non-negative integers $h$ we denote by $R(Q,h)$ the number of non-degenerate $h$-dimensional subspaces of $V$. A priori, this number depends on $Q$, but we will see that under our choice of $h$ it only depends on $k$ and $h$. In those cases we denote it by $R(k,h)$.

All lemmas above together prove the following.

\begin{Lemma}\label{lemma:construction1}
Let $h$ be chosen as above, assume that $R(k,h)=R(Q,h)$ is independent of the choice of a quadratic form $Q$.
Then
$$
N(k)=\frac{\gbinom{k}{h}q^{h(k-h)}}{R(k,h)}N(h)N(k-h).
$$
\end{Lemma}

\begin{Remark}
By classification theorems, any quadratic form can be obtained by Construction~\ref{construction 1} with $h=2$, independently of the rank parity. So it is natural to ask why, in the odd characteristic case, we are dealing separately with odd rank and even rank quadratic forms, using $h=1$ in the first case and $h=2$ in the second.
The reason is that if $\rk Q$ is odd then $R(Q,2)$ depends on $Q$, yielding a formula more complicated than the one given by Lemma~\ref{lemma:construction1}, involving terms which also depend on $Q$.
So our strategy allows a simpler proof.
\end{Remark}

Computing the number $R(k,h)$ is the last non trivial step towards the computation of $N(k)$.
We are going to do that in the next two sections, obtaining the following recursion formula.

\begin{Theorem}\label{thm:recursion-gen}
For $k\geq1$,
$$
N(k)=
\begin{cases}
(q^k-1)N(k-1)&\text{if }k\text{ is odd,}\\
q^kN(k-1)&\text{if }k\text{ is even.}
\end{cases}
$$
\end{Theorem}

Theorem~\ref{thm:recursion-gen} will be proved in the next two sections, dealing with the odd characteristic case and with the characteristic $2$ case separately. We now use it to prove the closed-form expression for $N(k)$ stated by Theorem~\ref{thm:fullrank-gen}.
Then we will conclude this section with the proof of Theorem~\ref{thm:givenrank-gen}.

\begin{Proof}[ of Theorem~\ref{thm:fullrank-gen}]
We argue by induction on $k$. First note that $N(0)=1$ and $N(1)=q-1$. Now let $k>1$ and assume that the statement is true for $k-1$. We use the recursion formula given by Theorem~\ref{thm:recursion-gen}. If $k$ is odd then
\begin{align*}
N(k)&=(q^k-1)N(k-1)=\\
&=(q^k-1)q^{\frac{k-1}{2}\left(\frac{k-1}{2}+1\right)}\prod_{i=1}^\frac{k-1}{2}(q^{2i-1}-1)=\\
&=q^{\frac{k-1}{2}\frac{k+1}{2}}\prod_{i=1}^\frac{k+1}{2}(q^{2i-1}-1).
\end{align*}
If $k$ is even then
\begin{align*}
N(k)&=q^kN(k-1)=\\
&=q^kq^{\frac{k}{2}\left(\frac{k}{2}-1\right)}\prod_{i=1}^\frac{k}{2}(q^{2i-1}-1)=\\
&=q^{\frac{k}{2}\left(\frac{k}{2}+1\right)}\prod_{i=1}^\frac{k}{2}(q^{2i-1}-1).
\end{align*}
\end{Proof}


\begin{Proof}[ of Theorem~\ref{thm:givenrank-gen}]
Consider the following construction.
For any choice of a subspace $V_0$ of dimension $r$, a full-rank quadratic form $Q_0$ on $V_0$ and a direct complement $R$ of $V_0$ we can define the quadratic form $Q:=Q_{(V_0,Q_0,R)}:=Q_0\oplus0\in\Quad(V)$ of rank $r$, i.e.\ the unique quadratic form on $V$ defined by the conditions ${Q|}_{V_0}=Q_0,{Q|}_R=0$ and $V_0\perp R$. By classification of quadratic forms, any rank $r$ quadratic form is given by $Q_{(V_0,Q_0,R)}$ for some triple $(V_0,Q_0,R)$.

So we only need to compute the number of times each form is repeated, i.e.\ the number of triples $(V_0',Q_0',R')$ such that $Q_{(V_0',Q_0',R')}=Q_{(V_0,Q_0,R)}=:Q$, where $(V_0,Q_0,R)$ is a fixed triple. First note that
$$
R'=\left\{x\in\Rad V:Q(x)=0\right\}=R,
$$
hence $V_0'$ has to be a direct complement of $R$.
But for any direct complement $V_0'$ of $R$ we have that the triple $(V_0',{Q|}_{V_0'},R)$ defines the form $Q$.
So, for any triple $(V_0,Q_0,R)$, the number of triples $(V_0',Q_0',R')$ such that $Q_{(V_0',Q_0',R')}=Q_{(V_0,Q_0,R)}$ is equal to the number of direct complements of $R$.

We are ready to conclude. We have $\gbinom{k}{r}$ choices for $V_0$, $N(r)$ choices for $Q_0$ by definition, $q^{r(k-r)}$ choices for $R$ by Lemma~\ref{lemma:0} and any form occurs $q^{r(k-r)}$ times. Hence $N(k,r)=\gbinom{k}{r}N(r)$, as claimed.
\end{Proof}

The next two sections constitute the proof of Theorem~\ref{thm:recursion-gen}. They share a similar structure: first we compute $R(k,h)$ in some interesting cases, then we use it, together with Lemma~\ref{lemma:construction1}, to prove Theorem~\ref{thm:recursion-gen}. Section~\ref{subsec:odd} deals with the odd characteristic case, Section~\ref{subsec:2} deals with the characteristic $2$ case.

\subsubsection{Odd characteristic case}\label{subsec:odd}

{\em In this section, assume that $\charf\Fq$ is odd.}

\begin{Lemma}\label{lemma:repetitions-odd}
We have that
\begin{enumerate}
\item $R(k,1)=q^{k-1}$ if $k$ is odd,
\item $R(k,2)=q^{k-2}\frac{q^k-1}{q^2-1}$ if $k$ is even.
\end{enumerate}
In particular, these numbers are independent of the choice of a full-rank quadratic form $Q$.
\end{Lemma}
\begin{Proof}
Let $Q$ be a full-rank quadratic form on $V$.
All $1$-dimensional subspaces $V_1\leq V$ such that ${Q|}_{V_1}$ has full rank are given by $V_1=\langle v_1\rangle$ for some vector $v_1\in V$ such that $Q(v_1)\not=0$. As $Q$ has odd rank, it has $q^{k-1}$ zeros, hence we have $q^k-q^{k-1}$ possible choices for $v_1$. But $\langle\lambda v_1\rangle=\langle v_1\rangle$ for any $\lambda\in\Fq,\lambda\not=0$, hence each subspace is counted $q-1$ times. So $R(k,1)=\frac{q^k-q^{k-1}}{q-1}=q^{k-1}$, and this proves the first claim.

We now prove the second claim. We can choose any non zero $v_1\in V$ as first basis vector of $V_1$ and we want to count the number of vectors $v_2\in V\setminus\langle v\rangle$ such that ${Q|}_{\langle v_1,v_2\rangle}$ has full rank. This holds if and only if
$$
\det\begin{pmatrix}
\tilde B_Q(v_1,v_1)&\tilde B_Q(v_1,v_2)\\
\tilde B_Q(v_1,v_2)&\tilde B_Q(v_2,v_2)
\end{pmatrix}
\not=0,
$$
i.e.\ if and only if $v_2$ is not a zero of the quadratic form on $V$ defined by
$$
Q'(x):=\tilde B_Q(v_1,v_1)\tilde B_Q(x,x)-{\tilde B_Q(v_1,x)}^2
$$
for $x\in V$. One can easily verify that this is indeed a quadratic form and that the associated bilinear form is defined by
$$
\tilde B_{Q'}(x,y)=2\tilde B_Q(v_1,v_1)\tilde B_Q(x,y)-2\tilde B_Q(v_1,x)\tilde B_Q(v_1,y)
$$
for $x,y\in V$.
We distinguish two cases.
If $\tilde B_Q(v_1,v_1)=0$ then $Q'(x)=-{\tilde B_Q(v_1,x)}^2$ is the square of a non zero linear form, hence it has rank $1$.
If $\tilde B_Q(v_1,v_1)\not=0$ then the radical of $V$ with respect to $\tilde B_{Q'}$ is exactly the span of $v_1$, hence $\rk Q'=\rk Q-1$ is odd as $\rk Q$ is even.
In order to prove this, let $w\in\Rad V$ (with respect to $\tilde B_{Q'}$), i.e.\ $\tilde B_{Q'}(w,y)=0$ for all $y\in V$. Then
\begin{align*}
\tilde B_{Q'}(w,y)&=2\tilde B_Q(v_1,v_1)\tilde B_Q(w,y)-2\tilde B_Q(v_1,w)\tilde B_Q(v_1,y)=\\
&=2\tilde B_Q(\tilde B_Q(v_1,v_1)w-\tilde B_Q(v_1,w)v_1,y)=0
\end{align*}
for all $y\in V$. But $\tilde B_Q$ is non-degenerate, hence this implies that $\tilde B_Q(v_1,v_1)w=\tilde B_Q(v_1,w)v_1$, therefore $w\in\langle v_1\rangle$ as $\tilde B_Q(v_1,v_1)\not=0$. This proves that $\Rad V\subseteq\langle v_1\rangle$, and the converse inclusion is obvious.
So in any case $\rk Q'$ is odd, hence $Q'$ has $q^{k-1}$ zeros.
We can finally conclude. We have $q^k-1$ choices for $v_1$ and $q^{k}-q^{k-1}$ choices for $v_2$, and any subspace is given by $(q^2-1)(q^2-q)$ different choices of $v_1,v_2$ (corresponding to the number of bases of $\langle v_1,v_2\rangle$). So we have $R(k,2)=\frac{(q^k-1)(q^{k}-q^{k-1})}{(q^2-1)(q^2-q)}=q^{k-2}\frac{q^k-1}{q^2-1}$. This concludes the proof.
\end{Proof}

The following theorem implies Theorem~\ref{thm:recursion-gen} in the odd characteristic case. First we need two remarks. Full-rank quadratic forms on $\Fq$ correspond to non zero elements of $\Fq$, hence $N(1)=q-1$. Full-rank quadratic forms on $\Fq^2$ correspond to triples $(x,y,z)\subseteq\Fq^3$ such that $xy-z^2\not=0$, which is a quadratic form of rank $3$, hence $N(2)=q^3-q^2=q^2(q-1)$.

\begin{Theorem}
For $k\geq 1$,
$$
N(k)=
\begin{cases}
(q^k-1)N(k-1)&\text{if }k\text{ is odd,}\\
q^k(q^{k-1}-1)N(k-2)&\text{if }k\text{ is even.}
\end{cases}
$$
\end{Theorem}
\begin{Proof}
If $k$ is odd then we apply Construction~\ref{construction 1} with $h=1$.
By Lemma~\ref{lemma:construction1} and the first claim of Lemma~\ref{lemma:repetitions-odd} we have
\begin{align*}
N(k)&=\frac{\gbinom{k}{1}q^{k-1}}{R(k,1)}N(1)N(k-1)=\\
&=\frac{q^k-1}{q-1}\frac{q^{k-1}}{q^{k-1}}(q-1)N(k-1)=\\
&=(q^k-1)N(k-1).
\end{align*}

If $k$ is even then we apply Construction~\ref{construction 1} with $h=2$.
By Lemma~\ref{lemma:construction1} and the second claim of Lemma~\ref{lemma:repetitions-odd} we have
\begin{align*}
N(k)&=\frac{\gbinom{k}{2}q^{2(k-2)}}{R(k,2)}N(2)N(k-2)=\\
&=\frac{(q^k-1)(q^{k-1}-1)}{(q^2-1)(q-1)}q^{2(k-2)}\times\\
&\times\frac{1}{q^{k-2}}\frac{q^2-1}{q^k-1}q^2(q-1)N(k-2)=\\
&=q^k(q^{k-1}-1)N(k-2).
\end{align*}
\end{Proof}

\subsubsection{Characteristic $2$ case}\label{subsec:2}

{\em In this section, assume that $\charf\Fq=2$.}

\begin{Lemma}\label{lemma:repetitions-2}
We have that
\begin{enumerate}
\item $R(k,2)=q^{k-2}\frac{q^k-q}{q^2-1}$ if $k$ is odd,
\item $R(k,2)=q^{k-2}\frac{q^k-1}{q^2-1}$ if $k$ is even.
\end{enumerate}
In particular, these numbers are independent of the choice of a full-rank quadratic form $Q$.
\end{Lemma}
\begin{Proof}
The proof is similar to the proof of the second claim of Lemma~\ref{lemma:repetitions-odd}.
Let $Q$ be a full-rank quadratic form on $V$.
In order to obtain a plane $\langle v_1,v_2\rangle\leq V$ such that ${Q|}_{\langle v_1,v_2\rangle}$ has full rank, we can choose any $v_1\in V\setminus\Rad V$ and any $v_2\in V\setminus\langle v_1\rangle$ which is not a zero of the quadratic form defined by
$$
Q'(x):=\tilde B_Q(v_1,v_1)\tilde B_Q(x,x)-{\tilde B_Q(v_1,x)}^2={\tilde B_Q(v_1,x)}^2
$$
for $x\in V$.
In the characteristic $2$ case this form always has rank $1$, hence it has $q^{k-1}$ zeros.
So we have $q^k-|\Rad V|$ choices for $v_1$ and $q^k-q^{k-1}$ choices for $v_2$, and any subspace is given by $(q^2-1)(q^2-q)$ different choices of $v_1,v_2$, hence $R(k,2)=\frac{(q^k-|\Rad V|)(q^{k}-q^{k-1})}{(q^2-1)(q^2-q)}=q^{k-2}\frac{q^k-|\Rad V|}{q^2-1}$. Now note that $|\Rad V|=q$ if $k$ is odd and $|\Rad V|=1$ if $k$ is even, hence both claims follow at once.
\end{Proof}

We are going to conclude the proof of Theorem~\ref{thm:recursion-gen}. Again, we use the fact that $N(2)=q^2(q-1)$.


\begin{Theorem}
For $k\geq1$,
$$
N(k)=\begin{cases}
q^{k-1}(q^k-1)N(k-2)&\text{if }k\text{ is odd,}\\
q^k(q^{k-1}-1)N(k-2)&\text{if }k\text{ is even.}
\end{cases}
$$
\end{Theorem}
\begin{Proof}
Recall that in this case we use Construction~\ref{construction 1} with $h=2$.
By Lemma~\ref{lemma:construction1} we have
\begin{align*}
N(k)&=\frac{\gbinom{k}{2}q^{2(k-2)}}{R(k,2)}N(2)N(k-2)=\\
&=\frac{1}{R(k,2)}q^{2(k-2)}q^2(q-1)\times\\
&\times\frac{(q^k-1)(q^{k-1}-1)}{(q^2-1)(q-1)}N(k-2).
\end{align*}
If $k$ is odd then by claim {\em 1} of Lemma~\ref{lemma:repetitions-2} we have
\begin{align*}
N(k)&=\frac{q^2-1}{q^k-q}\frac{1}{q^{k-2}}q^{2(k-2)}q^2(q-1)\times\\
&\times\frac{(q^k-1)(q^{k-1}-1)}{(q^2-1)(q-1)}N(k-2)=\\
&=q^{k-1}(q^k-1)N(k-2).
\end{align*}
If $k$ is even then by claim {\em 2} of Lemma~\ref{lemma:repetitions-2} we have
\begin{align*}
N(k)&=\frac{q^2-1}{q^k-1}\frac{1}{q^{k-2}}q^{2(k-2)}q^2(q-1)\times\\
&\times\frac{(q^k-1)(q^{k-1}-1)}{(q^2-1)(q-1)}N(k-2)=\\
&=q^k(q^{k-1}-1)N(k-2).
\end{align*}
\end{Proof}

%
%
%
%

\section*{Acknowledgment}

We wish to thank an anonymous reviewer for valuable comments that helped substantially improve the paper.



%




\end{document}